\documentclass[twocolumn,showpacs,showkeys,amsmath,amssymb]{revtex4}

\usepackage{graphicx}

\begin{document}

\title{Kerr black hole lensing for generic observers in the strong deflection limit}

\author{V. Bozza$^{a,b}$, F. De Luca$^{a,b,c}$, G. Scarpetta$^{a,b,d}$}

\affiliation{$^a$ Dipartimento di Fisica ``E.R. Caianiello'',
Universit\`a di Salerno, via Allende,
I-84081 Baronissi (SA), Italy.\\
  $^b$  Istituto Nazionale di Fisica Nucleare, Sezione di
 Napoli. \\
 $^c$ Institut f\"{u}r Theoretische Physik der
           Universit\"{a}t Z\"{u}rich, CH-8057 Z\"{u}rich, Switzerland \\
 $^d$ International Institute for Advanced Scientific Studies, Vietri sul Mare (SA),
Italy.}

\date{\today}

\begin{abstract}
We generalize our previous work on gravitational lensing by a Kerr
black hole in the strong deflection limit, removing the
restriction to observers on the equatorial plane. Starting from
the Schwarzschild solution and adding corrections up to the second
order in the black hole spin, we perform a complete analytical
study of the lens equation for relativistic images created by
photons passing very close to a Kerr black hole. We find out that,
to the lowest order, all observables (including shape and shift of
the black hole shadow, caustic drift and size, images position and
magnification) depend on the projection of the spin on a plane
orthogonal to the line of sight. In order to break the degeneracy
between the black hole spin and its inclination relative to the
observer, it is necessary to push the expansion to higher orders.
In terms of future VLBI observations, this implies that very
accurate measures are needed to determine these two parameters
separately.
\end{abstract}

\pacs{95.30.Sf, 04.70.Bw, 98.62.Sb}

\keywords{Relativity and gravitation; Classical black holes;
Gravitational lensing}

\maketitle

\section{Introduction}

As predicted by General Relativity, photons passing near a black
hole suffer deviations from their original trajectory. If the
minimum distance between photon and black hole is much larger than
the gravitational radius, a weak field approximation of the metric
tensor is sufficient to describe the light deflection. Two images
of the original source are then detected by the observer. On the
other hand, photons passing very close to the black hole may
suffer very large deviations without falling into the black hole.
These photons may perform one or more loops around the black hole
before reemerging in the observer direction, thus generating two
infinite sets of relativistic images very close to the black hole
shadow. It can be easily intuited that these relativistic images
represent a unique probe to gain information on the very strong
gravitational fields surrounding the black holes. Through their
study it would be possible to learn the properties of black holes
and get new insight on General Relativity in a strong field
regime. The features of relativistic images will thus represent a
possible challenge arena for alternative theories of gravitation.

Even though a general relativity approach to this subject
typically results in involved equations and heavy numerical
integrations, a surprisingly simple formula for the deflection
angle induced by a Schwarzschild black hole in the Strong
Deflection Limit (SDL) was proposed by Darwin \cite{Dar} and
revived in Refs. \cite{Lum,Oha,BCIS}. The logarithmic divergence
of the deflection angle in the impact parameter was showed to be
universal and valid for all spherically symmetric black holes
\cite{Boz1}, as every class of such black holes leads to the same
expansion for the deflection angle, with coefficients depending on
the specific form of the black hole metric. The SDL method was
then applied to several classes of black holes, ranging from
Reissner-Nordstr\"om to black holes in string theory, from
braneworld black holes to wormholes \cite{Spheric}. By the SDL
method it is thus possible to quantify the observables related to
relativistic images for any class of spherically symmetric black
holes, allowing an easy comparison among different theories. For
alternative methods, see Refs. \cite{Alternative,VirEll}.

For spinning black holes, things do not work so easily. Starting
from the geodesics equations in Kerr spacetime, that Carter
\cite{Car} reduced to first order equations depending on four
constants of motion, many numerical approaches have been developed
to study and visualize such geodesics. Numerical efforts have also
been profused in the context of gravitational lensing to
investigate the apparent shape of the accretion disk of the black
hole \cite{Lum,FMA,Accret}, the light curve of a star orbiting
around it \cite{CunBar} and the structure of the caustics
\cite{RauBla}, which turned out to be extended and to have a
4-cusped astroid structure. Some interesting general results have
recently been derived through Morse theory \cite{HasPer}. The
extension of the SDL methodology to Kerr black holes was firstly
performed in Ref. \cite{BozEq} and the SDL formula was recovered
for photons lying near to the equatorial plane. Anyway the
expansion coefficients had to be calculated numerically as
functions of the lens spin.

A first step toward a complete analytical treatment of this
subject was made in Ref. \cite{BDLSS} (hereafter Paper I) where
the lens equation was analytically solved in the limit of small
values of the angular momentum of the black hole (denoted by $a$)
and for observers lying on its equatorial plane. This last
assumption, besides ensuring simpler equations, was justified by
the fact that the most important black hole candidate (Sgr A*,
firstly suggested in Ref. \cite{VirEll}) is hosted in the center
of our Galaxy and presumably has a spin-axis perpendicular to the
Galactic plane, where the solar system lies. The expansions for
small values of the angular momentum allowed to use the
Schwarzschild SDL formula as a starting point for the description
of the deflection of light rays looping around a Kerr black hole.

This analytical approach provided very simple equations (which
could even be inverted for sources near to a caustic) and a full
description of the extended structure of the caustics, which were
confirmed to have a 4-cusped structure, symmetric w.r.t. the
equatorial plane and shifted from the optical axis. Only the first
order caustic cannot be recovered in the SDL approach as it is
formed in the weak deflection regime \cite{RauBla,Ser}. It was
also shown that the extension of relativistic caustics enhances
the cross section for the creation of additional images, whose
magnification is sensible in a relatively large region around the
caustic. Direct observations of these relativistic images, which
should be possible with the resolutions achieved by future
projects, could test the Kerr nature of black hole lenses (see
e.g. Refs. \cite{S1-S14,BDLSS} for detailed discussions on
observational perspectives).  It is interesting to compare the
lensing effect of a Kerr black hole to that of a Schwarzschild
black hole embedded in an external gravitational field. Also in
the latter case astroid caustics arise, though with different
sizes and positions \cite{BozSer}.

In this Paper we further investigate Kerr black hole lensing,
getting rid of the {\it equatorial observer} hypothesis. In spite
of the presence of a new parameter (the inclination of the spin
axis relative to the line of sight, that we shall indicate by
$\vartheta_o$), the surprisingly simple structure of all
analytical results is preserved. Our philosophy will be to try to
confine all technicalities to the appendices or refer the reader
to Paper I for more detailed derivations. This paper will thus
keep its main focus on the implications of all results on
observable quantities. What emerges from our study is that all
observables (to the lowest order) just depend on $a\sin
\vartheta_o$, that is the projection of the spin on a plane
orthogonal to the line of sight. The consequences of this fact
will be discussed in the conclusions in Section 7.

Our paper is structured as follows: in Section 2 we recall the
main properties of Kerr geodesics. In Section 3, we trace the
borders of the shadow of the Kerr black hole for all values of the
observer declination. In Section 4 we apply the SDL to null Kerr
geodesics illustrating the main strategy and referring to two
appendices for the details. In Section 5 we derive the critical
curves and caustics structure and in Section 6 we analyze the lens
mapping in the neighbourhood of a caustic, finding the position
and the magnification of the images, concluding with a discussion
on the detectability of relativistic images.

\section{Kerr geodesics}

In this section, we shall review the basics of Kerr geodesics and
introduce the notations to be used throughout the paper. For more
details on the physical meaning of all quantities, the reader may
refer to Paper I or directly to Ref. \cite{Cha}.

The main subject of our paper is the Kerr black hole, whose metric
in Boyer-Lindquist coordinates \cite{BoyLin}, $x^\mu \equiv
(t,x,\vartheta,\phi)$  reads

\begin{eqnarray}
& ds^2=&\frac{\Delta-a^2 \sin^2 \vartheta}{\rho^2}d
t^2-\frac{\rho^2}{\Delta} dx^2-\rho^2 d\vartheta^2 \nonumber \\
&& - \frac{ \left(x^2+a^2 \right)^2 - a^2\Delta \sin^2 \vartheta
}{\rho^2} \sin^2 \vartheta d\phi^2 \nonumber \\&&+\frac{2ax
\sin^2\vartheta}{\rho^2} dt d\phi \\%
& \Delta=&x^2-x+a^2, \\%
& \rho^2=& x^2+a^2 \cos^2\vartheta.
\end{eqnarray}
Distances are measured in units of the Schwarzschild radius
($2MG/c^2=1$), $\vartheta$ and $\phi$ are the colatitude and
azimuth respectively, $x$ is the radial coordinate and $a$ is the
specific angular momentum of the black hole, running from $0$
(Schwarzschild black hole) to $1/2$ (extremal Kerr black hole) in
our units.

\begin{figure}
\resizebox{\hsize}{!}{\includegraphics{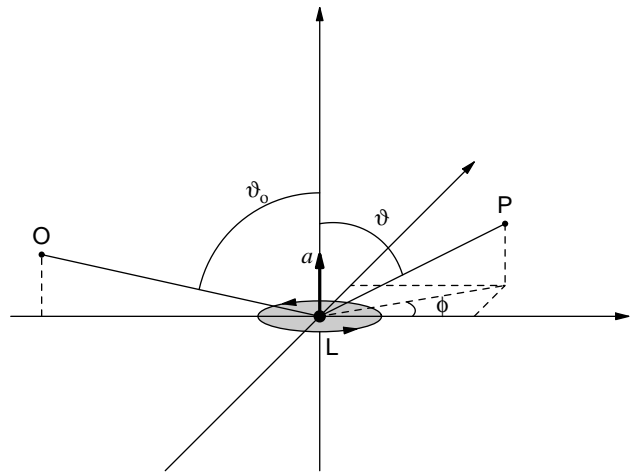}}
\caption{Boyer-Lindquist coordinates in a Kerr metric, also
referred as spin-oriented coordinates in the text. $L$ is the
black hole with spin $a$. $O$ is the observer and $P$ is a generic
point. The gray disk visualizes the equatorial plane of the black
hole. }
 \label{Fig SpinOri}
\end{figure}

We consider a static observer at Boyer-Lindquist coordinates
$(D_{OL},\vartheta_o,\phi_o)$. The distance between observer and
black hole is thus $D_{OL}$, while the colatitude $\vartheta_o$ of
the observer coincides with the inclination of the spin on the
line of sight $\overline{OL}$. Exploiting the freedom to choose
the zero of the azimuth, we set $\phi_o=\pi$. We will very often
use the notation $\mu \equiv \cos \vartheta$. Thus we also define
$\mu_o \equiv\cos \vartheta_o$. Fig. \ref{Fig SpinOri} illustrates
Boyer-Lindquist coordinates for a generic point $P$ and for the
observer $O$ in particular.

Lightlike geodesics can be expressed in the following form in
terms of  the first integrals of motion $J$ and $Q$ found by
Carter \cite{Car}
\begin{eqnarray}
&& \pm \int \frac{dx}{\sqrt{R}}=\pm \int \frac{d
\vartheta}{\sqrt{\Theta}} \label{Geod1}\\
& \phi_f-\phi_i =& a \int\frac{x^{2}+a^{2}-a J}{\Delta \sqrt{R}}
dx-a \int \frac{dx}{\sqrt{R}} \nonumber  \\
&& + J \int \frac{\csc^2\vartheta}{\sqrt{\Theta}} d \vartheta,
\label{Geod2}
\end{eqnarray}
where
\begin{eqnarray}
&\Theta=&Q+a^2 \cos^2\vartheta-J^2 \cot^2\vartheta \\
&R=&x^4+(a^2-J^2-Q)x^2+(Q+(J-a)^2) x \nonumber \\ &&-a^2 Q,
\label{R}
\end{eqnarray}
and $\phi_i$ is the initial value of the azimuthal coordinate of
the photon.

The roots of $R$ represent inversion points in the radial motion.
In gravitational lensing we consider photons coming from infinity,
grazing the black hole and going back to infinity. For such
trajectories there is only one inversion point $x_0$, representing
the closest approach distance. The minimum allowed value of $x_0$
can be found solving the equations $R(x)=0$ and $R'(x)=0$
simultaneously. However, in Kerr black hole, we do not have a
unique minimum closest approach $x_m$, but rather a continuous
family of values which depend on the approach trajectory followed
by the photon. In particular, it is possible to establish a
relation among the minimum closest approach $x_m$ and the
corresponding values of the constants of motion $J$ and $Q$, that
we shall indicate by $J_m$ and $Q_m$ (see e.g. Ref. \cite{Cha})
\begin{eqnarray}
&&J_{m}=\frac{x_{m}^{2}(2 x_{m}-3)+a^{2}(1+2 x_{m})}{a(1-2
x_{m})} \label{Jm}\\
&& Q_{m}=\frac{x_{m}^{3}\left[ 2
a^2-x_{m}(x_{m}-3/2)^2\right]}{a^{2}(x_{m}-1/2)^2}. \label{Qm}
\end{eqnarray}
$x_m$ also represents the radius of the unstable circular photon
orbit. This radius is fixed to $3/2$ when $a=0$ (Schwarzschild
black hole). In the case of Kerr black holes, $x_m$ may vary
between two limiting values $x_{m+}$, $x_{m-}$, depending on the
incoming direction of the photon. The two limiting values can be
analytically obtained solving the equation $Q_m=0$ (in fact, it is
possible to show that gravitational lensing trajectories cannot be
realized for $Q<0$ \cite{Cha}). To the third order in $a$, they
read \cite{BDLSS}
\begin{equation}
x_{m\pm}=\frac{3}{2}\mp \frac{2}{\sqrt{3}} a -\frac{4}{9} a^2 \mp
\frac{20}{27\sqrt{3}}a^3+ O(a^4). \label{xmpm}
\end{equation}

For example, photons whose orbit lies on the equatorial plane may
turn either in the same sense of the black hole (prograde photons)
or in the opposite sense (retrograde photons). Prograde photons
are allowed to get closer to the black hole, with a minimum
closest approach given by $x_{m+}$, while retrograde photons must
stay farther than $x_{m-}$, in order to be deflected without
falling into the black hole. Photons whose orbit does not lie on
the equatorial plane are characterized by intermediate values of
$x_m$, with $Q_m>0$. Thus $x_m$ can be used to parametrize the
family of unstable photon orbits allowed in Kerr metric or,
equivalently, the incoming direction of the photon. The
corresponding values of the constants of motion are uniquely
determined by Eqs. (\ref{Jm}) and (\ref{Qm}).

Although exact expressions for $x_{m+}$ and $x_{m-}$ are
available, it is convenient to start with a perturbative expansion
{\it ab initio} in order to be prepared to face more complicated
quantities in the following \cite{BDLSS}. Throughout our
treatment, only for $x_m$ we need to push the expansion to the
third order, in order to obtain some quantities to the second
order in $a$.

\section{The shadow of a Kerr black hole}

\begin{figure}
\resizebox{\hsize}{!}{\includegraphics{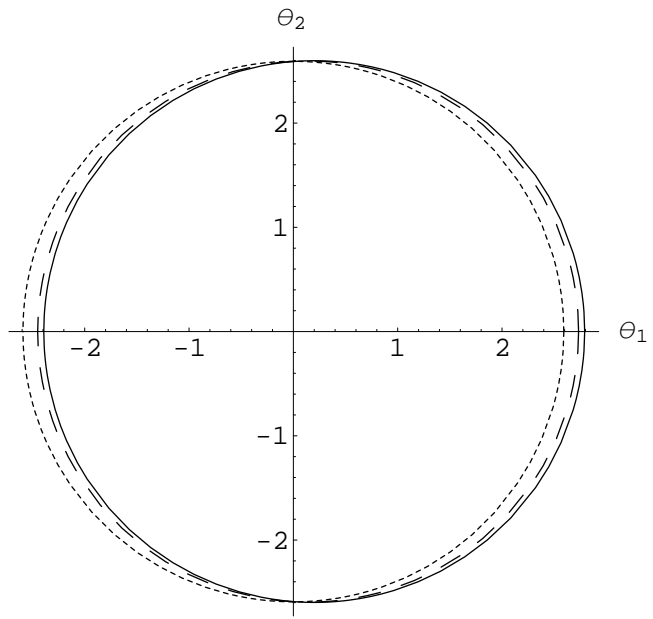}} \caption{ The
shadow of the black hole in the observer sky for $a=0.1$ and
different values of the observer position $\vartheta_o$. The solid
line is for $\vartheta_o=\pi/2$ (equatorial observer), the dashed
line is for $\vartheta_o=\pi/4$ and the dotted line is for
$\vartheta_o=0$ (polar observer).}
 \label{Fig Thm}
\end{figure}

The constants of motion $J$ and $Q$ have an immediate link to the
position in the sky where the observer detects the photon. In
fact, we can define angular coordinates $(\theta_1,\theta_2)$ on
the observer sky centered on the black hole position. We choose
the orientation of these coordinates in such a way that the spin
axis of the black hole is projected on the $\theta_2$-axis (see
Fig.\ref{Fig Thm}).

As shown in Ref. \cite{Cha}, photons detected by the observer at
angular coordinates $(\theta_1,\theta_2)$ are characterized by
constants of motion given by
\begin{eqnarray}
&& J=-\theta_{1} D_{OL} \sqrt{1-\mu_o^2}, \label{JTh1} \\
&& Q=\theta_{2}^2 D_{OL}^2 +\mu_o^2(\theta_{1}^2
D_{OL}^2-a^2).\label{QTh2}
\end{eqnarray}
These relations can be easily recovered taking the limit for large
distances in the equations of motion of the photon. They show that
$J$ can be identified with the component of the orbital angular
momentum of the photon along the spin axis, whereas $Q+J^2+\mu_o^2
a^2$ is the squared total angular momentum of the photon.

Note that, with our choice of $(\theta_1,\theta_2)$, in the limit
of equatorial observer $\mu_o=0$, prograde photons ($J>0$, $Q=0$)
are detected by the observer on the left side of the black hole,
while retrograde photons ($J<0$, $Q=0$) are detected on the right
side. Conversely, in the limit of polar observers ($\mu_o
\rightarrow \pm 1$), the projected angular momentum $J$ vanishes,
while $Q \rightarrow (\theta_{1}^2+\theta_{2}^2) D_{OL}^2 -a^2$.

Inverting Eqs. (\ref{JTh1}) and (\ref{QTh2}), we find the position
$(\theta_1,\theta_2)$ in the sky where the photon is detected with
given constants of motion $J$ and $Q$, apart from an ambiguity in
the sign of $\theta_2$
\begin{eqnarray}
&& \theta_{1}=- \frac{J}{D_{OL} \sqrt{1-\mu_o^2}}, \label{Th1J} \\
&& \theta_{2}=\pm D_{OL}^{-1}\sqrt{Q + \mu_o^2\left( a^2-
\frac{J^2}{1-\mu_o^2} \right) }.\label{Th2Q}
\end{eqnarray}

These relations can be used to convert the locus $(J_m,Q_m)$,
parametrized by $x_m$ according to Eqs. (\ref{Jm}) and (\ref{Qm})
in the $(J,Q)$-space, into a new one $(\theta_{1,m},\theta_{2,m})$
in the observer sky. However, not all values of $x_m$ in the range
$[x_{m+},x_{m-}]$ are acceptable. This can be easily understood,
as photons lying on the equatorial plane can never reach
non-equatorial observers. The reality condition for $\theta_{2,m}$
restricts $x_m$ to the range $[x_{p+},x_{p-}]$, where $x_{p+}$ and
$x_{p-}$ are the roots of the equation $\theta_{2,m}=0$. To third
order in $a$, these quantities read
\begin{eqnarray}
&x_{p\pm}=& \frac{3}{2}\mp\frac{2}{\sqrt{3}}a
\sqrt{1-\mu_o^2}-\frac{4}{9}a^2(1+\mu_o^2) \nonumber \\ &&
\mp\frac{4}{27\sqrt{3}} a^3(5+6\mu_o^2)\sqrt{1-\mu_o^2}+O(a^4).
\label{xp}
\end{eqnarray}
Comparing with Eq. (\ref{xmpm}), we see that $x_{p\pm} \rightarrow
x_{m\pm}$ in the limit $\mu_o \rightarrow 0$. On the other hand,
when $\mu_o \rightarrow \pm 1$, the allowed range for $x_m$
shrinks to a single value $ x_{p} \rightarrow
\frac{3}{2}-\frac{8}{9}a^2$. This witnesses that when the observer
is on the polar axis the axial symmetry of the lensing
configuration is restored and all unstable photon orbits have the
same radius again.

When $a$ vanishes, $x_{p+}$ and $x_{p-}$ both coincide with the
Schwarzschild photon sphere radius, $3/2$, while, when $a$ is not
zero, they are distinct and every value of $x_m$ in the interval
$[x_{p+},x_{p-}]$ uniquely fixes the amplitude of the oscillation
of the photon orbit on the equatorial plane, as we shall see
later. On the basis of this consideration, in Paper I (with
$\mu_o=0$) we introduced a parametrization of $x_m$ in the range
$[x_{m+},x_{m-}]$, replacing $a$ with $a \xi$ in Eq. (\ref{xmpm}),
with the parameter $\xi$ varying in the range $[-1,1]$.

In order to take into account the changes from Eq. (\ref{xmpm}) to
(\ref{xp}), we have to update such parametrization, since it is
not directly applicable to the case $\mu_o \neq 0$. Our new
parametrization for $x_m$ is
\begin{eqnarray}
&x_{m}=& \frac{3}{2}-\frac{2}{\sqrt{3}}a\xi
\sqrt{1-\mu_o^2}-\frac{4}{9}a^2(1+\mu_o^2) \nonumber \\ &&
-\frac{4}{27\sqrt{3}} a^3\xi(5+6\mu_o^2)\sqrt{1-\mu_o^2}+O(a^4).
\label{xmxi}
\end{eqnarray}

As $\xi$ varies in the interval  $[-1,1]$ we get all possible
values of $x_m$ in the interval $[x_{p+},x_{p-}]$. It will become
clear later that $\xi$ is strictly related to the position angle
of the generic point in the observer sky.

With this parametrization, we can rewrite
Eqs.(\ref{Jm})-(\ref{Qm}) to the second order in $a$ as
\begin{eqnarray}
&J_m(\xi)=&\frac{3\sqrt{3}}{2}\xi\sqrt{1-\mu_o^2}-a(1-\mu_o^2)(1+\xi^2)
\nonumber
\\ && -a^2 \frac{\xi \sqrt{1-\mu_o^2}}{3\sqrt{3}}[5-2 \xi^2-2
\mu_o^2(1-\xi^2)], \label{Jma2}
\end{eqnarray}
\begin{eqnarray}
&Q_m(\xi)=&\frac{27}{4}\left[ 1-(1-\mu_o^2)\xi^2 \right] \nonumber \\
&&
-3\sqrt{3}a\xi \sqrt{1-\mu_o^2}[1+\mu_o^2-(1-\mu_o^2)\xi^2] \nonumber \\
&& -a^2[(1+\mu_o^2)^2-4(1-\mu_o^2)\xi^2 \nonumber \\
&& +3(1-\mu_o^2)^2\xi^4]. \label{Qma2}
\end{eqnarray}
Notice that the presence of $a$ in the denominators of
Eqs.(\ref{Jm})-(\ref{Qm}) allows $\xi$ to be present already in
the zero-order terms in Eqs.(\ref{Jma2})-(\ref{Qma2}), permitting
the use of the $\xi$-parametrization in Schwarzschild spacetime as
well. However, since this parametrization has been introduced in a
slightly different way w.r.t. Paper I, the expressions derived
here cannot be directly compared to those of Paper I, except for
those quantities that are independent of $\xi$. For example,
eliminating $\xi$ from Eqs. (\ref{Jma2}) and (\ref{Qma2}), one can
derive an expression for the locus $(J_m,Q_m)$ in the form
$Q_m(J_m)$. Doing the same with the expressions of Paper I, one
would indeed find the same function $Q_m(J_m)$ in the limit $\mu_o
\rightarrow 0$.

Inserting Eqs. (\ref{Jma2}) and (\ref{Qma2}) in Eqs.
(\ref{Th1J})-(\ref{Th2Q}) we get
\begin{eqnarray}
&D_{OL} \theta_{1,m}&=-\frac{3\sqrt{3}}{2}\xi+a
\sqrt{1-\mu_o^2}(1+\xi^2) \nonumber \\ &&
+a^2\frac{\xi}{3\sqrt{3}}[5-2\mu_o^2 -2 \xi^2(1-\mu_o^2)],
\label{Th1m}
\end{eqnarray}

\begin{eqnarray}
&D_{OL} \theta_{2,m}& = \pm\frac{3\sqrt{3}}{2}\sqrt{1-\xi^2} \mp
a\xi \sqrt{1-\xi^2} \sqrt{1-\mu_o^2} \nonumber \\ &&  \mp a^2
\frac{\sqrt{1-\xi^2}}{3\sqrt{3}}[1+2\mu_o^2-2\xi^2(1-\mu_o^2)].
\label{Th2m}
\end{eqnarray}
This locus is formed by the points in the observer sky where
photons with minimum closest approach would show up. No
gravitational lensing images are possible inside this locus, which
is thus also known as the shadow of the black hole. In Fig.
\ref{Fig Thm} we show it for different values of $\mu_o$. Note
that, to zero order, $\theta_{1,m} \propto -\xi$ and $\theta_{2,m}
\propto \sqrt{1-\xi^2}$, justifying the identification of $\xi$
with the cosine of the position angle in the $(\theta_1,\theta_2)$
plane as taken from the opposite of the $\theta_1$-axis. This fact
facilitates the physical interpretation of the parameter $\xi$.

The shadow of the black hole is the first observable in extreme
gravitational lensing by supermassive black holes. It thus
deserves some further analysis in order to understand the effect
of the spin and the observer position.

First we note that $\theta_{1,m}$ and $\theta_{2,m}$, to second
order in $a$, satisfy the ellipse equation
\begin{equation}
\frac{(\theta_{1,m}-\theta_0)^2}{A_1^2}+
\frac{\theta_{2,m}^2}{A_2^2}=1 \label{shadow}
\end{equation}
with the origin shifted rightward by
\begin{equation}
\theta_0=\frac{2a\sqrt{1-\mu_o^2}}{D_{OL}}, \label{ShadowShift}
\end{equation}
and semiaxes given by
\begin{eqnarray}
&& A_1=D_{OL}^{-1} \left(\frac{3\sqrt{3}}{2}-\frac{a^2}{\sqrt{3}} \right) \\
&& A_2=D_{OL}^{-1}
\left(\frac{3\sqrt{3}}{2}-\frac{a^2}{\sqrt{3}}\mu_o^2 \right).
\end{eqnarray}

By these analytical expressions for the shadow, we can make
several considerations. The presence of a non-vanishing spin
causes a slight distortion and a displacement of the shadow from
the black hole position. When the observer lies on the spin axis
($\mu_o=\pm 1$), the axial symmetry is restored and the shadow
returns to be centered on the black hole and circular. However,
even in this limiting case, the radius of the shadow is no longer
$3\sqrt{3}/2$ as in Schwarzschild, but it is slightly smaller,
being $3\sqrt{3}/2-a^2/\sqrt{3}$.

It has been proposed that the observation of the shape of the
shadow of a black hole by VLBI may help to determine the
parameters of a Kerr black hole, such as its mass, its angular
momentum and the inclination of the spin \cite{FMA,Zak}. However,
both in the shift $\theta_0$ and in the ellipticity
\begin{equation}
e\equiv\frac{A_2-A_1}{A_2}=\frac{2}{9}a^2(1-\mu_o^2)
\end{equation}
the black hole spin and the observer declination appear in the
same combination $a\sqrt{1-\mu_o^2}=a \sin \vartheta_o$, which
represents the projection of the spin on a plane orthogonal to the
line of sight. Thus it is impossible to determine both the
absolute value of the spin and its inclination from the shape of
the shadow. The only possibility is that we already know the
distance $D_{OL}$ and the mass of the black hole to such accuracy
that we are able to extract $a$ from a measure of the minor
semi-axis $A_1$ solely. However, since the spin contribution to
the major semi-axis is only of second order in $a$, we need a very
high accuracy in the shadow observation in order to appreciate
such a small contribution. For example, if $a=0.1$, the spin
contribution to $A_1$ is of order $0.2\%$. As already pointed in
Ref. \cite{Zak} by numerical examples, the disentanglement of $a$
and $\vartheta_o$ is only possible for values of the black hole
spin very close to the extremal case. By our perturbative
formulae, we have justified this claim analytically. Of course,
for high values of $a$, when higher orders contribute to determine
the shape of the shadow, the degeneracy between $a$ and
$\vartheta_o$ can be broken, in agreement with what stated in Ref.
\cite{Zak}.

It has been pointed out in Paper I that as long as we deal with
Kerr black holes with spin smaller than $a=0.2$, the perturbative
approximation works surprisingly well. Then, the degeneracy
between $a$ and $\vartheta_o$ in the shadow of the black hole
poses a serious difficulty to the determination of the parameters
of the black hole by the simple observation of the shadow. As we
shall see in the next sections, this degeneracy plagues all
gravitational lensing observables in different degrees.

\section{Kerr lensing in the Strong Deflection Limit}

As in Paper I, we introduce the following parametrization of the
observer sky
\begin{equation}
\left\{ \begin{array}{l}
 \theta_1(\epsilon,\xi)=\theta_{1,m}(\xi)(1+\epsilon) \\
\theta_2(\epsilon,\xi)=\theta_{2,m}(\xi)(1+\epsilon)
\end{array}
\right. . \label{ThetaParam}
\end{equation}
Varying $\xi$ in the range $[-1,1]$ and $\epsilon$ in the range
$[-1,\infty]$, we can obviously cover the whole upper half of the
observer sky, since $\xi$ establishes the position angle of the
light ray w.r.t. the $(-\theta_1)-$axis (through Eqs. (\ref{Th1m})
and (\ref{Th2m})) whereas $\epsilon$ fixes the angular distance
from the shadow of the black hole. In fact, in the following,
$\epsilon$ will be generically referred to as the separation from
the shadow of the black hole.

We are interested into light rays experiencing very large
deflections by a Kerr black hole. These light rays reach the
observer from directions $(\theta_1,\theta_2)$ very close to the
shadow. In the parametrization (\ref{ThetaParam}), they are thus
characterized by a very small positive $\epsilon$, while keeping
$\xi$ in the whole range $[-1,1]$. The SDL amounts to performing
the integrals in the geodesics equations
(\ref{Geod1})-(\ref{Geod2}), to the lowest orders in the
separation from the shadow $\epsilon$.

The values of the constants of motion $J$ and $Q$, corresponding
to such strongly deflected photons, can be found using Eqs.
(\ref{JTh1})-(\ref{QTh2}):
\begin{eqnarray}
&J(\xi,\epsilon)=& J_m(\xi) (1+\epsilon) \label{JSFL}\\
&Q(\xi,\epsilon)=& Q_m(\xi) (1+2\epsilon) +2 a^2 \epsilon \mu_o^2. \label{QSFL}
\end{eqnarray}

Substituting these expressions in Eq. (\ref{R}) and solving the
equation $R=0$ for $x_0$, we get the closest approach distance as
\begin{eqnarray}
&& \!\!\!\!\!\!\!\!\!\!\!\!\!\!\! x_0(\xi,\delta)= x_m(\xi)(1+\delta) \label{x0SFL} \\
&& \!\!\!\!\!\!\!\!\!\!\!\!\!\!\! \delta =
\sqrt{\frac{2\epsilon}{3}}\left[ 1 -\frac{2}{3\sqrt{3}}a\hat\xi
+\frac{2}{27}a^2 (10-\mu_o^2-14 \hat\xi^2)\right]
\label{deltatoeps}
\end{eqnarray}
where we have introduced the compact notation
\begin{equation}
\hat\xi=\xi \sqrt{1-\mu_o^2}. \label{xindef}
\end{equation}

As $\epsilon$ represents the separation of the image in the
observer sky from the shadow of the black hole, $\delta$
represents the relative difference between the closest approach
$x_0$ and the minimum closest approach $x_m(\xi)$ fixed by the
position angle through $\xi$. It will be synthetically called
approach parameter. As $\delta$ decreases, we expect the
deflection to increase more and more. In the limit $\delta
\rightarrow 0$, the photon is injected into the unstable orbit
with radius $x_m(\xi)$. Conversely, photons with a large approach
parameter are weakly deflected. Of course, the relation between
$\epsilon$ and $\delta$ ensures that the SDL can be equivalently
stated in terms of either of the two parameters.

Let us introduce our gravitational lensing configuration. As said
before, the observer is at radial coordinate $D_{OL}$, at polar
angle $\vartheta_o$ and azimuthal angle $\phi_o=\pi$. We will call
optical axis the line connecting the lens and the observer. The
source is assumed to be static at Boyer-Lindquist coordinates
$(D_{LS},\vartheta_s,\phi_s)$.

Our lens equations are provided by Eqs.
(\ref{Geod1})-(\ref{Geod2}), where we identify the final value of
the azimuthal coordinate with the observer's one
($\phi_f=\phi_o=\pi$), and the initial value with the source's one
$\phi_i=\phi_s$. In these equations there are two radial integrals
and two angular integrals. The radial integrals are solved using
the SDL technique and expanding all coefficients to second order
in $a$, as in Paper I. The results of this procedure are reported
in Appendix A. Similarly, the angular integrals are solved to
second order in $a$ in Appendix B.

Once all integrals are calculated, we have to solve Eqs.
(\ref{Geod1})-(\ref{Geod2}) in terms of the source coordinates
$(\phi_s,\mu_s)$, so that they are in the lens map form
\begin{equation}
\left\{
\begin{array}{l}
\mu_s=\mu_s(\delta,\xi)  \\
\phi_s=\phi_s(\delta,\xi)
\end{array} \right. . \label{LensApp}
\end{equation}

Note that the lens equation will be written in terms of the
approach parameter $\delta$ and the position angle through $\xi$.
Through Eqs. (\ref{deltatoeps}) and (\ref{ThetaParam}) we can then
go back to the coordinates in the observer sky
$(\theta_1,\theta_2)$.

In the following sections, we will calculate the critical curves
and the caustics of the Kerr gravitational lens order by order.
The procedure is indeed identical to that described in Paper I,
save for the complication introduced by the additional parameter
$\mu_o$. However, once we manage to recast all equations in the
best forms, the results remain very simple, so that a thorough
discussion of the effects of spin and observer colatitude is
possible.

%%%%%%%%%%%%%%%%%%%%%%%%%%%%%%%%%%%%%%%%%%%%%%%%%%%%%%%%%%%%%%%%%%
%%%%%%%%%%%%%%%%%%%%%%%%%%%%%%%%%%%%%%%%%%%%%%%%%%%%%%%%%%%%%%%%%%

\section{Derivation of the relativistic caustics}

\subsection{Zero-order caustics}
\label{ZeroOrder}

The first task is to recover the results for a Schwarzschild black
hole, imposing the limit $a\rightarrow 0$.

Using the results of Appendices A and B to the zero-order, Eqs.
(\ref{Geod1}) and (\ref{Geod2}) read respectively
\begin{eqnarray}
& \psi & = m\pi\mp \arcsin \frac{\mu_s}{\sqrt{1-{\hat\xi}^2}}
\nonumber \\ && \pm (-1)^m \arcsin
\frac{\mu_o}{\sqrt{1-{\hat\xi}^2}}  , \label{PreEqmu0} \\
& \phi_s & =  \pi-Sign[\xi]m\pi \pm\arctan \frac{\mu_s
{\hat\xi}}{\sqrt{1-\mu_s^2-{\hat\xi}^2}} \nonumber \\ && \mp(-1)^m
\arctan \frac{\mu_o {\hat\xi}}{\sqrt{1-\mu_o^2-{\hat\xi}^2}} ,
  \label{PreEqphi0}
\end{eqnarray}
where the new variable
\begin{equation}
\psi=-2\log \delta
+2\log[12(2-\sqrt{3})],\label{psidef}
\end{equation}
allows us to put the equations in a very compact form. $\psi$
actually coincides with the deflection induced by a Schwarzschild
black hole with the same mass of our Kerr black hole. On the
ground of this connection, we shall often refer to $\psi$ as
``scalar deflection'' in the following.

The double signs coming from the angular integrals must be treated
as follows: if the photon moves out of the source increasing its
initial value of $\mu$, then we must choose the upper signs,
otherwise we will select the lower signs. These double signs are
the relics of those present in Eqs. (\ref{Geod1}) and
(\ref{Geod2}). For more details about their origin, the reader is
referred to the Appendix B. $m$ represents the number of
inversions in the polar motion of the photon.

Introducing the quantity
\begin{equation}
\psi_o=\mp(-1)^m\arcsin\frac{\mu_o}{\sqrt{1-{\hat\xi}^2}},\label{psi0def}
\end{equation}
we can easily solve Eqs. (\ref{PreEqmu0})-(\ref{PreEqphi0}) w.r.t.
$\phi_s$ and $\mu_s$ to get the zero-order lens equation
\begin{eqnarray}
&\mu_s &= \mp(-1)^m \sqrt{1-{\hat\xi}^2}\sin( \psi+\psi_o), \label{Eqmu0} \\
& \phi_s &= \pi(1-m) +\arctan \left[\hat\xi \tan \psi_o \right]
\nonumber  \\  &&  -\arctan \left[\hat\xi \tan
(\psi+\psi_o)\right]. \label{Eqphi0}
\end{eqnarray}
Since the azimuth $\phi$ is a coordinate with period $2\pi$, we
have eliminated the $Sign[\xi]$ in front of $m\pi$ in Eq.
(\ref{Eqphi0}). In the derivation of Eq. (\ref{Eqphi0}) from Eqs.
(\ref{PreEqphi0}) and (\ref{Eqmu0}), we have used the relations
\begin{eqnarray}
&& \frac{\mu_s }{\sqrt{1-\mu_s^2-{\hat\xi}^2}} = \mp \tan (\psi+\psi_o) \\
&& \frac{\mu_o }{\sqrt{1-\mu_o^2-{\hat\xi}^2}} = \mp (-1)^m \tan
\psi_o
\end{eqnarray}
and exploited the fact that the number of inversions in the polar
motion is just the integer part of $(\psi+\psi_o+\pi/2)/\pi$.

Let us understand the meaning of the zero-order lens equations.
Eq. (\ref{Eqmu0}) states that the photon performs symmetric
oscillations on the equatorial plane (recall that $\mu\equiv \cos
\vartheta$) with amplitude $\sqrt{1-\hat\xi^2}$, which depends on
the observer declination and the trajectory chosen by the photon
(polar $\hat\xi=0$, equatorial $\hat\xi=\pm 1$ or whatever). The
number of oscillations depends on the scalar deflection $\psi$,
which diverges when the approach parameter $\delta \rightarrow 0$.
$\psi_o$ is the initial condition of the oscillation, which
depends on the observer declination. The double signs take into
account the fact that the oscillations occur in opposite ways
depending on the starting sign of $\dot\mu$.

Eq. (\ref{Eqmu0}) is the azimuthal motion of the photon. It can be
better understood when we choose equatorial photons with
$\hat\xi=1$. Then it just reduces to $\phi= \pi-\psi$, which
states that the azimuthal shift is the scalar deflection minus
$\pi$, as expected in this simple case. Different values of $\xi$
need to be analyzed by some spherical trigonometry, in order to
understand the trigonometric functions in Eq. (\ref{Eqmu0}).

After the zero order lens equation is constructed, we can study
the structure of critical curves and caustics. The Jacobian of the
lens map, $D$, can be easily calculated from (\ref{Eqmu0}) and
(\ref{Eqphi0}). We find
\begin{eqnarray}
&& \frac{\partial \mu_s}{\partial \xi}= \pm(-1)^m \frac{\hat\xi \sqrt{1-{\mu_o}^2}}{\sqrt{1-{\hat\xi}^2}} \sec \psi_o \sin \psi \\
&& \frac{\partial \mu_s}{\partial \psi}= \mp(-1)^m \sqrt{1-{\hat\xi}^2}\cos (\psi+\psi_o)\\
&& \frac{\partial \phi_s}{\partial \xi}= - \frac{
\cos(\psi+\psi_o) \sec\psi_o \sin \psi}{\sqrt{1-\mu_o^2}}
\\ && \frac{\partial \phi_s}{\partial
\psi}= -\frac{\hat\xi \sec^2 (\psi+\psi_o)}{1+{\hat\xi}^2 \tan^2
(\psi+\psi_o)}
\end{eqnarray}
and using Eqs. (\ref{xindef}) and (\ref{psi0def}), we finally have
\begin{equation}
D= \frac{\partial \mu_s}{\partial \xi} \frac{\partial
\phi_s}{\partial \psi}-\frac{\partial \mu_s}{\partial
\psi}\frac{\partial \phi_s}{\partial \xi}= \mp(-1)^m \frac{ \sin
\psi}{\sqrt{1-{\xi}^2}}.
\end{equation}

Since all transformations from $(\psi,\xi)$ to
$(\theta_1,\theta_2)$ are non-singular (except for the points
$\xi=\pm 1$), the solutions of the equation $D=0$ determine the
critical curves. To zero order we have
\begin{equation}
\psi_{k}= k \pi. \label{CC0}
\end{equation}

As expected, the critical curves correspond to values of the
scalar deflection that are multiples of $\pi$. Having introduced
the most generic coordinate system for the black hole has not
prevented us from recovering the Schwarzschild result. Through
Eqs. (\ref{psidef}), (\ref{deltatoeps}) and (\ref{ThetaParam}) we
reconstruct the critical curves in the observer coordinates
\begin{equation}
\begin{array}{l}
D_{OL} \theta_{1,k}(\xi)= -\frac{3 \sqrt{3} \xi}{2} \left[1+ \epsilon_{k} \right]\\ \\
D_{OL} \theta_{2,k}(\xi) = \pm  \frac{3 \sqrt{3}}{2}
\sqrt{1-\xi^2}\left[1+ \epsilon_{k} \right]
\end{array}, \label{Crit0}
\end{equation}
where

\begin{equation}
\epsilon_{k}=216(2-\sqrt{3})^2e^{-k\pi} \label{epscrit0}
\end{equation}
is the separation of the critical curve from the shadow.

We will refer to the integer number $k$ as the critical curve (or
caustic) order. Eqs. (\ref{Crit0}) describe a series of concentric
rings, parametrized by $\xi$, slightly larger than the shadow of
the black hole and whose radius
$\frac{3\sqrt{3}}{2}(1+\epsilon_k)$ exponentially decreases to the
shadow radius with increasing critical curve order.

The equations of the caustics are easily found introducing Eq.
(\ref{CC0}) into (\ref{Eqmu0})-(\ref{Eqphi0}) and exploiting the
fact that the number of inversions $m$ coincides with $k$ if
$\psi=k\pi$. We have
\begin{equation}
\mu_s=(-1)^k \mu_o, ~~ \phi_s=(1-k)\pi.
 \label{cau0}
\end{equation}

As already known, the Schwarzschild caustics are point-like and
lie on the optical axis. They are in front of the black hole
($\mu_s=\mu_o$, with $\phi_s$ being an odd multiple of $\pi$) for
even values of $k$ (retrolensing caustics), and behind it
($\mu_s=-\mu_o$, with $\phi_s$ being an even multiple of $\pi$)
for odd $k$ (standard lensing).

The SDL description is limited to large deflections ($\psi \gtrsim
\pi$), thus working better and better for higher order caustics
\cite{Boz1,S1-S14}. It cannot be applied to the first order one
($k=1$) whose full description can be derived in the weak
deflection limit for sources sufficiently far from the lens. In
what follows, we focus on caustics of order $k\geq2$ and
investigate how their structure is affected by the concomitant
action of the lens spin and the observer declination.

\subsection{First-order caustics}

We now introduce first order corrections to the zero-order
solutions found in the previous section. Starting from the results
of Appendix A and B, we solve the lens equations perturbatively
adding the first order terms to Eqs. (\ref{Eqmu0})-(\ref{Eqphi0}),
obtaining

\begin{eqnarray}
&\mu_s= &\mp(-1)^m \sqrt{1-{\hat\xi}^2}\sin(\psi+ \psi_o)
\nonumber
\\ && \mp(-1)^m \frac{2a \hat\xi}{3\sqrt{3}}\sqrt{1-{\hat\xi}^2}\cos \psi_o
\sin \psi,\label{Eqmu1}
\end{eqnarray}

\begin{eqnarray}
& \phi_s =& (1-m)\pi+\arctan \left[\hat\xi \tan \psi_o \right]
 \nonumber \\ &&-\arctan \left[\hat\xi \tan(\psi+\psi_o) \right]
\nonumber
\\ && -\frac{4a}{3\sqrt{3}}\left[\psi +3 \sqrt{3} \log(2
\sqrt{3}-3)\right.  \nonumber \\ && \left. -
\frac{1-{\hat\xi}^2}{2} \frac{ \cos(\psi+\psi_o) \sin \psi \cos
\psi_o}{1-(1-{\hat\xi}^2) \sin^2(\psi+\psi_o)} \right].
 \label{Eqphi1}
\end{eqnarray}

The Jacobian of the lens equation to first order is
\begin{equation}
D= \mp(-1)^m \frac{ \sin \psi}{\sqrt{1-{\xi}^2}}
\left(1+\frac{2a\xi\sqrt{1-\mu_o^2}}{\sqrt{3}} \right),
\end{equation}
which is always solved by Eq. (\ref{CC0}), thus implying that the
scalar deflection $\psi$ and consequently the approach parameter
$\delta$ are not affected by lens spinning to the first order.
Anyway, due to the spin dependence in Eq. (\ref{deltatoeps}),
first order corrections modify the separation of the critical
curves from the shadow. They read
\begin{eqnarray}
& D_{OL} \theta_{1,k}= &-\frac{3 \sqrt{3} \xi}{2}(1+\epsilon_{k})+ \nonumber \\ && a \left[1+\xi^2+\epsilon_{k}(1- \xi^2)\right]\sqrt{1-{\mu_o}^2}, \nonumber \\
& D_{OL} \theta_{2,k}= & \pm \frac{3
\sqrt{3}(1+\epsilon_{k})\sqrt{1-\xi^2}}{2}\mp \nonumber \\ && a
\xi \sqrt{1-\xi^2}(1-\epsilon_{k})\sqrt{1-{\mu_o}^2},
\label{Crit1}
\end{eqnarray}
where $\epsilon_k$ is still the zero-order separation defined in
Eq. (\ref{epscrit0}).

Coming to the caustics, from Eqs. (\ref{Eqmu1})-(\ref{Eqphi1}) and
(\ref{CC0}) we get
\begin{eqnarray}
&& \mu_s=(-1)^k \mu_o, \\ && \phi_s = \pi(1-k)-\Delta \phi_k \\
&& \Delta \phi_k= 4a\left[\frac{k\pi}{3\sqrt{3}} - \log(2
\sqrt{3}-3)\right]. \label{Cau1}
\end{eqnarray}

So, caustics are still point-like but the alignment with the
optical axis is now missing because of first order corrections, as
already pointed out in Paper I. The azimuthal shift is
proportional to the caustic order, it does not depend on the
observer declination and is negative, thus implying a clockwise
drift, if we look at the black hole from the north pole. This
means that, as $k$ is still the number of inversion points,
prograde (retrograde) light rays, emitted by a source on a caustic
point, perform more (respectively less) than $(k-1)/2$ loops.
Moreover, as the caustics drift from the optical axis and from
each other, perfect alignment of observer, lens and source is not
required for the enhancement of the images which are enhanced one
at a time, as sources cannot cross more than one caustic point at
the same time. For numerical values of the shift see Paper I,
Table 1.

\subsection{Second-Order Caustics}
  \label{cau2par}
In this section we investigate the effects of second order
corrections in the black hole spin on the critical curves and
caustics. Following the same steps as in the previous subsection,
we can add the second order terms $a^2\delta \mu_s^{(2)}$ to Eqs.
(\ref{Eqmu1}) and $a^2\delta \phi_s^{(2)}$ to (\ref{Eqphi1}).
Since they have quite long expressions, we report them in Appendix
C and proceed with the analysis of the second order lens equation.
In fact, although the general second order lens equation is quite
involved, it is easy to solve the Jacobian determinant equation
$D=0$ in terms of the second order perturbation of $\psi$,
starting from the zero order solution (\ref{CC0}). We get
\begin{equation}
\psi_{k}=k \pi+a^2 \delta\psi, \label{CCtot}
\end{equation}
where
\begin{eqnarray}
& \delta\psi=& -\frac{1}{18} \left[9c_k
(3-2\mu_o^2-3\hat\xi^2)+32(1-\hat\xi^2)\right] \label{psi_2}
\end{eqnarray}
and
\begin{equation}
c_k=\frac{2}{9}(5k\pi+8\sqrt{3}-36). \label{xc}
\end{equation}

Using Eqs. (\ref{psidef}) and (\ref{deltatoeps}) we can calculate
the second order corrections to the approach parameter $\delta$
and the separation from the shadow $\epsilon$. After that, by Eqs.
(\ref{ThetaParam}), we can derive the second order corrections to
the critical curves given in Eq. (\ref{Crit1})
\begin{equation}
\begin{array} {lll}
D_{OL} \theta_{1,k}^{(2)} & = & a^2 \frac{\xi}{3
\sqrt{3}}\left[5-2\mu_o^2-2 \hat\xi^2 \right. \\ &&
\left. +\epsilon_{k}\left(29-8\mu_o^2-32 \hat\xi^2\right) \right],\\ \\
D_{OL} \theta_{2,k}^{(2)} & = & \mp a^2 \frac{\sqrt{1-\xi^2}}{3
\sqrt{3}}\left[1+2\mu_o^2-2\hat\xi^2 \right.\\ && \left.
+\epsilon_{k}\left(21-32 \hat\xi^2\right)\right], \label{Crit2}
\end{array}
\end{equation}
where the zero-order separation $\epsilon_{k}$ is always given by
Eq. (\ref{epscrit0}).

Plugging Eq. (\ref{CCtot}) into the lens map, we get the caustics
parametric equations up to the second order in $a$:
\begin{eqnarray}
&& \mu_s= (-1)^k\mu_o \pm a^2
c_k(1-{\mu_o}^2)^{3/2}(1-\xi^2)^{3/2},
\label{Caumu2} \\
&& \phi_s=(1-k)\pi-\Delta \phi_k-a^2 c_k \xi^3\sqrt{1-{\mu_o}^2}.
\label{Caugamma2}
\end{eqnarray}
As explained in Section \ref{ZeroOrder}, the double sign in Eq.
(\ref{Caumu2}) allows for the possibilities that the photon starts
its journey by increasing $\mu$ or by decreasing $\mu$,
respectively. it is necessary to take both possibilities into
account in order to cover the whole caustic. In agreement with
Paper I and other works where the same results are found
numerically (e.g.\cite{RauBla}), we get extended caustics whose
shape is a 4-cusped astroid, with cusps in $\xi=\pm1$ and $\xi=0$
(for different signs of initial $\dot \mu$). The extension of the
caustics along $\mu$ and along $\phi$ is different. However,
choosing appropriate coordinates centered on the caustic, it is
possible to show that the extension in the sky as seen by the
black hole is the same along both axes (see next subsection).

\subsection{Observables related to critical curves and caustics}

After second order corrections to critical curves and caustics
have been derived, we can discuss their dependence on $a$ and
$\vartheta_o$.

First we note that the critical curves obtained adding Eq.
(\ref{Crit2}) to (\ref{Crit1}) satisfy the ellipse equation
\begin{equation}
\frac{(\theta_{1,k}-\theta_0)^2}{A_{1,k}^2}+
\frac{\theta_{2,k}^2}{A_{1,k}^2}=1 \label{CritEllipse}
\end{equation}
with the same origin shift as the shadow (Eq. (\ref{ShadowShift}))
and semiaxes given by
\begin{eqnarray}
& A_{1,k}&=D_{OL}^{-1} \left\{\frac{3\sqrt{3}}{2}(1+\epsilon_{k})
\right. \nonumber \\&& \left.
-a^2\frac{4- \epsilon_{k}(4-9 c_k \mu_o^2 )}{4\sqrt{3}}\right\} \\
& A_{2,k}&=D_{OL}^{-1} \left\{\frac{3\sqrt{3}}{2}- \left[
16\epsilon_{k}^2 -4(3+\epsilon_{k}^2)\mu_o^2\right.\right.
\nonumber \\ && \left.\left.
 \!\!\!\!\!\!\!\!\!\!\! +27c_k \epsilon_{k}
(1+\epsilon_{k})(3-2\mu_o^2)
\right]\frac{a^2}{12\sqrt{3}(1+\epsilon_{k})} \right\}.
\end{eqnarray}
The critical curves tend to coincide with the shadow in the limit
$k\rightarrow \infty$, which corresponds to photons winding an
infinite number of times, thus tending to the unstable photon
orbit. The ellipticity of the critical curves is
\begin{equation}
e=a^2(1-\mu_o^2) \frac{4(3+\epsilon_{k}^2)+81c_k
 \epsilon_{k} (1+\epsilon_{k})}{54(1+\epsilon_{k})^2},
\end{equation}
which is slightly higher than that of the shadow for the lower
order critical curves, but tends to that of the shadow as
$k\rightarrow \infty$. In particular, we see that shift and
ellipticity of the critical curves still depend on the combination
$a\sin \vartheta_o$, as for the shadow. So, even the observation
of several critical curves cannot help to determine $a$ and
$\vartheta_o$ separately.

Let us come to the caustics. Here the situation is more subtle and
needs to be investigated with grain of salt.

Suppose we have no independent knowledge of the direction of the
black hole spin or, at least, the direction of the spin is not
known to any great accuracy. Then, the observer will construct his
coordinates allowing for a non-vanishing position angle $\nu$ for
the spin axis. The uncertainty in $\nu$ will be determinant in the
following discussion. Let us thus introduce $(x,\hat\vartheta,\hat
\phi)$ as observer-oriented coordinates, still centered at the
black hole, but with the polar axis perpendicular to the optical
axis and the azimuth $\hat\phi$ taken from the direction opposite
to the observer. In general, if the observer ignores the spin
axis, the spin axis of the black hole would have a position angle
$\nu$ from the polar axis as fixed by the observer. The coordinate
transformation from $(\mu,\phi)$ to $(\hat\vartheta,\hat \phi)$ is
\begin{eqnarray}
&\hat \vartheta & = \arccos \left[\mu\sqrt{1-\mu_o^2}\cos \nu
+\mu_o\sqrt{1-\mu^2}\cos \phi \cos \nu \right.
\nonumber \\ && \left.+ \sqrt{1-\mu^2}\sin \phi \sin \nu\right]  \\%
&\hat \phi &=\arctan \left[\left(\sqrt{1-\mu^2}\sin \phi \cos \nu
 \right.\right. \nonumber
\\ && \left. -  \mu_o\sqrt{1-\mu^2}\cos \phi \sin \nu - \mu \sqrt{1-\mu_o^2} \sin \nu
\right) \nonumber
\\&& \left. \cdot \left(\sqrt{1-\mu^2}\sqrt{1-\mu_o^2}\cos \phi-\mu \mu_o
\right)^{-1}  \right].
\end{eqnarray}

\begin{figure}
\resizebox{\hsize}{!}{\includegraphics{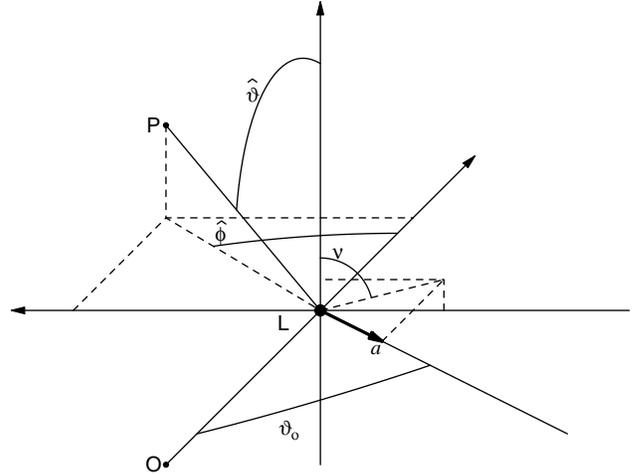}}
\caption{Observer-oriented coordinates $(\hat\vartheta,\hat\phi)$
introduced in the text. $L$ is the black hole with spin $a$. $O$
is the observer and $P$ is a generic point. $\vartheta_o$ is the
inclination of the spin on the line of sight, $\nu$ is the
position angle of the spin. }
 \label{Fig ObsOri}
\end{figure}

Fig. \ref{Fig ObsOri} illustrates the geometrical meaning of these
coordinates.

Transforming the caustics (\ref{Caumu2})-(\ref{Caugamma2}) from
the spin-oriented coordinates $(\mu,\phi)$ to the
observer-oriented coordinates $(\hat\vartheta,\hat \phi)$, and
expanding to second order in $a$, we get
\begin{eqnarray}
& \hat\vartheta_s &= \frac{\pi}{2}-(-1)^k \Delta \phi_k
\sqrt{1-\mu_o^2} \sin \nu \nonumber \\ && -(-1)^k\frac{1}{2}\Delta
\phi_k^2\mu_o
\sqrt{1-\mu_o^2}\cos \nu \nonumber \\ && \!\!\!\!\!\!\!\!\!\! - R_k\left[ (-1)^k\xi^3 \sin \nu \pm (1-\xi)^{3/2} \cos \nu \right] \label{cauhattheta}\\
& \hat \phi_s &= (1-k)\pi-\Delta \phi_k \sqrt{1-\mu_o^2} \cos \nu
\nonumber \\ && + \frac{1}{2}\Delta \phi_k^2\mu_o
\sqrt{1-\mu_o^2}\sin \nu \nonumber
\\ && \!\!\!\!\!\!\!\!\!\! -  R_k\left[ \xi^3 \cos \nu \pm (-1)^k (1-\xi)^{3/2} \sin \nu
\right], \label{cauhatphi}
\end{eqnarray}
where
\begin{equation}
R_k\equiv a^2c_k(1-\mu_o^2)=\frac{2}{9}a^2 (1-\mu_o^2)
(5k\pi+8\sqrt{3}-36) \label{deltacau}
\end{equation}
is the semi-amplitude of the caustic. In fact, we can appreciate
that, in observer-oriented coordinates, the extension of the
caustic is the same in both polar and azimuthal directions, as
anticipated before for any coordinate system centered on the
caustic. So, the extension is quadratic in the spin and is maximal
for equatorial observers, while the astroid shrinks to a single
point when the observer lies on the spin axis. The caustic
extension also increases linearly with the caustic order $k$.

Then, we note that the angular shift of the center of the caustic
from the optical axis is
\begin{eqnarray}
&\Delta_k&\equiv \arccos[\sin \hat\vartheta \cos \hat \phi] =
\Delta \phi_k \sqrt{1-\mu_o^2} \nonumber \\ && = 4a
\sqrt{1-\mu_o^2}\left[\frac{k\pi}{3\sqrt{3}} - \log(2
\sqrt{3}-3)\right]. \label{shift}
\end{eqnarray}
It is linear in the black hole spin and the caustic order.
Similarly to the semi-amplitude, also the shift is maximal for
equatorial observers and vanishes for polar observers, when the
axial symmetry is restored.

The shift and the semi-amplitude of the caustics are very easy
quantities to determine in case of observation of the relativistic
images generated by a source crossing a relativistic caustic. In
fact, if the observer is able to identify the source and follow
its direct image throughout the duration of the caustic crossing
event, then he would immediately determine the position of the
caustic and estimate its extension. Unfortunately, even in these
two quantities, the black hole spin and the observer declination
always appear in the combination $a\sqrt{1-\mu_o^2}=a \sin
\vartheta_o$, making the breaking of the degeneracy between these
two parameters impossible. On the other hand, it is easy to
determine the order $k$ of the caustic involved in the lensing
event, since the ratio
\begin{equation}
\frac{\Delta_k^2}{R_k}= \frac{8\left[k \pi+3\sqrt{3} \log
(2\sqrt{3}-3) \right]^2}{3\left( 5k\pi+8\sqrt{3}-36 \right)}
\end{equation}
only depends on $k$ and increases monotonically in $k$, without
degeneracy between any two values.

One possibility for the separate determination of $a$ and $\mu_o$
arises in case the spin position angle $\nu$ is known to a very
good accuracy from independent measures. Then we can move to a
more convenient coordinate frame where $\nu=0$. If this is
possible, looking at Eqs. (\ref{cauhattheta}) and
(\ref{cauhatphi}) we see that the shift in the azimuthal direction
is linear in $a$, while a residual quadratic shift is present in
the polar direction, which amounts to
\begin{eqnarray}
&\delta_k & \equiv \frac{1}{2} \Delta \phi_k^2 \mu_o
\sqrt{1-\mu_o^2} \nonumber \\ && =8a^2 \mu_o
\sqrt{1-\mu_o^2}\left[\frac{k\pi}{3\sqrt{3}} - \log(2
\sqrt{3}-3)\right]^2.
\end{eqnarray}

Then, if one is able to measure this residual shift, one can
extract the observer colatitude $\vartheta_o$ as
\begin{equation}
\cot \vartheta_o = \frac{2\delta_k}{\Delta_k^2}.
\end{equation}

Once the observer position relative to the spin axis is known, we
can use either $\Delta_k$ or $R_k$ to extract the black hole spin
$a$. However, as for the case of the direct determination of $a$
from the measure of the minor semi-axis of the shadow, this is a
higher order measure, which requires very accurate independent
information.

\begin{figure}
\resizebox{\hsize}{!}{\includegraphics{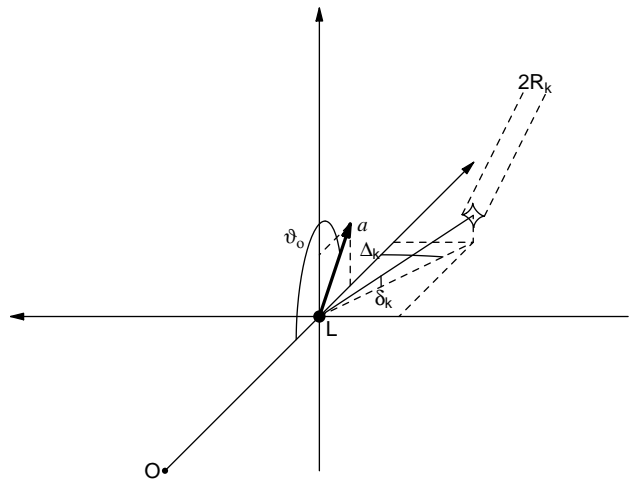}} \caption{A
typical caustic in Kerr lensing. The extension is the same in both
directions. Having chosen coordinates such that the position angle
of the spin vanishes, the caustic has an azimuthal shift
$\Delta_k$ and a vertical shift $\delta_k$ w.r.t. the line of
sight. }
 \label{Fig cau}
\end{figure}

Fig. \ref{Fig cau} shows a caustic and illustrates the meaning of
the semi-amplitude $R_k$, the horizontal shift $\Delta_k$ and the
vertical shift $\delta_k$. The picture is done for a standard
lensing caustic ($k$ odd) with $\vartheta_o>\pi/2$, so that the
caustic is displaced upward (see Eq. (\ref{cauhattheta})).

As usual we can trust our results as long as the perturbative
terms remain small. In extremal or close-to-extremal Kerr black
holes, higher orders in $a$ would play a major role in the
critical curves and caustics profile. In that case, the degeneracy
between $a$ and $\vartheta_o$ can be probably broken also through
the determination of the extension and position of the caustics or
through the analysis of the critical curves. However, in the
literature there is no investigation of Kerr black holes with high
spin that is deep enough to allow a comparison with our
perturbative results for low spins.

\section{Gravitational lensing near caustics}

\subsection{Position of the relativistic images}

Although in our picture the images cannot be found analytically
for arbitrary source positions using the lens mapping that we have
derived, they can be actually found for sources in the
neighbourhood of a caustic. This is indeed the most interesting
case, as the relativistic images are highly magnified and become
observable only if this event occurs. Assuming that the angular
distance between the source and a caustic of order $k$ is of the
order of $a^2$ (thus comparable with the caustic semiaxis), we can
write the source position as
\begin{equation}
\mu_s=(-1)^k \mu_o+a^2 \delta \mu_s, \label{mu_cau}
\end{equation}
\begin{equation}
\phi=(1-k)\pi-\Delta\phi_k+a^2 \delta \phi_s , \label{gamma_cau}
\end{equation}

In this assumption, the images will be very close to the critical
curve of order $k$. Then the scalar deflection will be
\begin{equation}
\psi=k \pi+a^2 \delta \psi.
\end{equation}

Plugging the last equation into the lens map written up to
corrections of second order in $a$ and inverting with respect to
$\delta \mu_s$ and $\delta \phi_s$, we get
\begin{eqnarray}
& \delta \mu_s= & \mp \frac{1}{9} \sqrt{1-\mu_o^2}
\sqrt{1-\xi^2}\left[9 \delta \psi \right.\nonumber \\ && \left.
+(5 k \pi+8 \sqrt{3}-20)(1-(1-{\mu_o}^2) \xi^2)\right],
\label{lensmu}
\end{eqnarray}

\begin{eqnarray}
&\delta \phi_s= & \frac{\xi}{9 \sqrt{1-{\mu_o}^2}}\left[92-24
\sqrt{3}-15 k \pi -9 \delta \psi \right. \nonumber \\ && +2
{\mu_o}^2
(5 k \pi+8 \sqrt{3}-36)   \nonumber \\
&&\left. +(5 k \pi+8 \sqrt{3}-20)(1-{\mu_o}^2) \xi^2\right].
\label{lensgamma}
\end{eqnarray}

Solving (\ref{lensgamma}) with respect to $\delta \psi$
%we get
%\begin{eqnarray}
%&\delta \psi =& -\frac{\delta \gamma \sqrt{1-{\mu_o}^2}}{\xi}+\frac{1}{9}(92-24 \sqrt{3}-15 k \pi \nonumber \\ &&+ {\mu_o}^2 (-72+16 \sqrt{3}+10 k \pi)
%\nonumber \\ && -(20-8 \sqrt{3}-5 k \pi)(1-{\mu_o}^2) \xi^2).
%\end{eqnarray}
%This is the expression for $\delta \psi$ that we insert in (\ref{lensmu}), so finding
and plugging its expression into (\ref{lensmu}), we find
\begin{equation}
\delta \mu_s \xi=-S (-1)^k(1-{\mu_o}^2)( \delta \phi_s+c_k
\sqrt{1-{\mu_o}^2} \xi) \sqrt{1-\xi^2}, \label{PreEqimages}
\end{equation}
where $c_k$ is given by Eq. (\ref{xc}) and $S=\mp (-1)^k$. This
equation can be more conveniently written in terms of
observer-oriented coordinates $(\hat \vartheta_s, \hat \phi_s)$.
Supposing that the position angle of the spin has been well
established by observations of the shadow or by the shift of the
caustic itself, we put $\nu=0$ for simplicity and write
\begin{equation}
\hat \vartheta_s=\frac{\pi}{2} - (-1)^k \delta_k +\delta \hat
\vartheta_s \label{hattheta_cau}
\end{equation}
\begin{equation}
\hat \phi_s=(1-k)\pi-\Delta\phi_k+\delta \hat \phi_s ,
\label{hatphi_cau}
\end{equation}
with
\begin{eqnarray}
&& \delta \hat \vartheta_s = -a^2 \frac{\delta
\mu_s}{\sqrt{1-\mu_o^2}} \label{deltahattheta} \\
&& \delta \hat \phi_s = a^2 \delta \phi_s \sqrt{1-\mu_o^2}.
\label{deltahatphi}
\end{eqnarray}

Then, we can write Eq. (\ref{PreEqimages}) directly in terms of
these coordinates as
\begin{equation}
\delta \hat \vartheta_s \xi = S(-1)^k (\delta \hat \phi_s +R_k
\xi) \sqrt{1-\xi^2}, \label{Eqimages}
\end{equation}
where $R_k$ is the semi-amplitude of the caustic given by Eq.
(\ref{deltacau}). The solutions of this equation for arbitrary
source positions $(\delta\hat \vartheta_s,\delta \hat\phi_s)$
determine the relativistic images generated by the Kerr black
hole. As the roots of Eq. (\ref{Eqimages}) are found squaring both
its sides, the solutions of the squared equation satisfy the
original one only for one choice of $S$. $S$ is directly related
to the half-sky where the image appears. In fact, we recall that
the parameterization (\ref{ThetaParam}) has an ambiguity in the
sign of $\theta_2$. This ambiguity can be solved observing that
the photon reaches the observer from the upper side of the black
hole if $S$ is positive and from the lower side if $S$ is
negative. This fact can be easily established remembering that in
all our equations the upper signs hold when the photon leaves the
source by increasing its $\mu$ coordinate. Then, if its polar
motion undergoes one inversion ($k=1$), the photon reaches the
observer from above and we coherently have $S=1$. On the other
hand, if the lower signs hold, the photon begins its motion
decreasing its $\mu$ coordinate. With one inversion, it reaches
the observer from below and coherently we have $S=-1$. The same
reasoning can be repeated with an arbitrary number $k$ of
inversions in the polar motion.

It can be easily verified that Eq. (\ref{Eqimages}) has four real
solutions if the source is inside the caustic and only two real
solutions if the source is outside. Once the coordinate $\xi$
(which, we recall, represents the cosine of the position angle) of
the image is known, Eq. (\ref{lensgamma}) can be used to
determines the value of $\delta \psi$ (perturbation of the scalar
deflection). However, it is important to stress that Eq.
(\ref{Eqimages}) determines $\xi$ to zero order only. Therefore,
though the positions of the images in the observer sky are
generically given by
\begin{eqnarray}
&D_{OL} \theta_1=& -\frac{3 \sqrt{3}}{2} \xi(1+\epsilon_{k})
\nonumber \\ && +a
\sqrt{1-{\mu_o}^2}\left[1+\xi^2+\epsilon_{k}(1-\xi^2)\right]+\nonumber \\
&& \frac{a^2 \xi}{ 6\sqrt{3}}\left\{10-4 \xi^2(1-\mu_o^2)-4
\mu_o^2+\epsilon_{k}\left[27 \delta \psi  \right.\right. \nonumber \\
&&\left.\left. +58-64 \xi^2(1-\mu_o^2)-16\mu_o^2\right] \right\},
\label{image1}
\end{eqnarray}

\begin{eqnarray}
&D_{OL} \theta_2=& S \frac{3 \sqrt{3}}{2}
\sqrt{1-\xi^2}(1+\epsilon_{k}) \nonumber \\ &&
-a S\xi\sqrt{1-\xi^2}\sqrt{1-\mu_o^2}(1-\epsilon_{k}) \nonumber \\
&& -S \frac{a^2}{6 \sqrt{3}}\sqrt{1-\xi^2}\left\{ 2-4
\xi^2(1-\mu_o^2)+4 \mu_o^2 \right. \nonumber \\
&& \left. +\epsilon_{k}\left[ 27\delta \psi+42-64(1-\mu_o^2)
\xi^2\right] \right\}, \label{image2}
\end{eqnarray}
to the second order in $a$, only a zero order expression of $\xi$
is actually available. So, the position of the images is accurate
only to zero order in $a$ and is given by
\begin{eqnarray}
&D_{OL} \theta_1=& -\frac{3 \sqrt{3}}{2} \xi(1+\epsilon_{k})
\label{2image1} \\ &D_{OL} \theta_2=& S \frac{3 \sqrt{3}}{2}
\sqrt{1-\xi^2}(1+\epsilon_{k}). \label{2image2}
\end{eqnarray}

To zero order, we see that the images of order $k$ lie along the
critical curve of order $k$ (we remind that $\epsilon_k$ is just
the separation of the critical curve of order $k$ from the shadow
(\ref{epscrit0})), with position angle determined by the solutions
of Eq. (\ref{Eqimages}). If a more accurate theoretical prediction
of the images position (including first order corrections) is
needed, it is necessary to push the lens equation to the third
order. Indeed this would be a worthy (though heavy) task since the
equation for the images (\ref{Eqimages}) depends on $a$ only
through $R_k$. As noticed before, this quantity only depends on
the projection of the spin on the line of sight. So, once more,
the observables (in this case the positions of the images) only
depend on $a\sin\vartheta_o$ to the lowest order. However,
contrarily to the former observables, the positions of the images
could be detected to an accuracy sufficiently high to be sensitive
at least to first order corrections in $a$. So, it would be indeed
desirable to check whether the positions of the images may help to
break the degeneracy between the absolute value of the spin and
its inclination on the optical axis.

\subsection{Magnification}

The magnification is defined as the ratio of the angular area of
the image and the corresponding angular area of the source. The
angular area of the image is simply $|d\theta_1 d\theta_2|$, while
the angular area of the source is $|\sin \vartheta_s d\phi_s
d\vartheta_s|$ or $|\sin \hat\vartheta_s d\hat\phi_s
d\hat\vartheta_s|$ if one uses observer-oriented coordinates. Then
the magnification can be calculated as $|\sin
\hat\vartheta_s|^{-1}$ times the inverse of the Jacobian
determinant of the lens application in the form

\begin{equation}
\left\{
\begin{array}{l}
\hat\phi_s=\hat\phi_s(\theta_1,\theta_2) \\
\hat\vartheta_s=\hat \vartheta_s(\theta_1,\theta_2)
\end{array}
\right. . \label{IdealLens}
\end{equation}

Following the same approach of Paper I, we can find the expression
of the magnification for sources in the neighbourhood of caustics
exploiting the available relations
(\ref{deltahatphi})-(\ref{deltahattheta}) and
(\ref{lensmu})-(\ref{lensgamma}) to get
\begin{equation}
\left\{
\begin{array}{l}
\delta \hat\phi_s = \delta \hat\phi_s (\delta \psi,\xi) \\
\delta \hat\vartheta_s = \delta \hat\vartheta_s (\delta \psi,\xi)
\end{array}
\right.
\end{equation}
and (\ref{image1})-(\ref{image2})
\begin{equation}
\left\{
\begin{array}{lll}
\theta_1 & = & \theta_1 (\delta \psi,\xi) \\
\theta_2 & = & \theta_2 (\delta \psi,\xi)
\end{array}
\right. .
\end{equation}
Then the perturbation of the scalar deflection $\delta \psi$ and
the cosine of the position angle $\xi$ play the role of
intermediate variables between the source coordinates
$(\hat\vartheta_s,\hat\phi_s)$ and the image coordinates in the
observer sky $(\theta_1,\theta_2)$.

Since $\sin\hat \vartheta_s d\hat\phi_s=d(\delta\hat\phi_s)$ and $
d\hat \vartheta_s=d(\delta\hat\vartheta_s)$ to the lowest order,
the Jacobian of the map (\ref{IdealLens}) reduces to
\begin{equation}
\frac{\partial (\hat\phi_s,\hat \vartheta_s)}{\partial
(\theta_1,\theta_2)} = \frac{\partial (\delta\hat\phi_s,\delta
\hat \vartheta_s)}{\partial (\delta \psi,\xi)}  \left[
\frac{\partial (\theta_1,\theta_2)} {\partial (\delta \psi,\xi)}
\right]^{-1},
\end{equation}
where we have used the matrix notation
\begin{equation}
\frac{\partial (y_1,y_2)}{\partial (x_1,x_2)}=
\left(
\begin{array}{cc}
\frac{\partial y_1}{\partial x_1} & \frac{\partial y_1}{\partial
x_2} \\
\frac{\partial y_2}{\partial x_1} & \frac{\partial y_2}{\partial
x_2}
\end{array}
\right).
\end{equation}

As the derivatives and the Jacobian matrix have very involved
expressions, we do not go too much into detail and only report
here the two eigenvalues of the Jacobian matrix
\begin{equation}
\lambda_r=\frac{2 D_{OL}}{3 \sqrt{3} \epsilon_{k}},
\label{eigenrad}
\end{equation}

\begin{equation}
\lambda_t=\frac{2 D_{OL} D_O}{27 \sqrt{3} (1+\epsilon_{k})},
\label{eigentan}
\end{equation}

where
\begin{eqnarray}
&D_O=& (-1)^k \frac{a^2}{2}
\left\{9c_k\left[3-2\mu_o^2-3\hat\xi^2\right] \right. \nonumber
\\ && \left. +32(1-\hat\xi^2) + 18 \delta \psi\right\}. \label{D_O}
\end{eqnarray}

In a first approximation $\lambda_r$ only depends on the caustic
order $k$ and is always positive. On the other hand $\lambda_t$
vanishes at caustic crossing (see. Eq.(\ref{psi_2})). Following
Paper I, we will call $\lambda_r$ and $\lambda_t$, respectively,
radial and tangential eigenvalues, although they are such only in
the limit $a \rightarrow 0$. Taking into account that the flux
received by the observer is $D_{LS}^2/D_{OS}^2$ times the flux
received by the black hole, the radial and tangential
magnifications are

\begin{equation}
\mu_r=\frac{D_{OS}}{D_{LS}} \frac{1}{\lambda_r},  \label{magnrad} \\
\end{equation}
\begin{equation}
\mu_t= \frac{D_{OS}}{D_{LS}} \frac{1}{|\lambda_t|}  \label{magntan} \\
\end{equation}
while the total magnification is given by $\mu=\mu_r \mu_t$.

An interesting thing to note is that the radial magnification is
completely independent of $a$ and $\mu_o$. It is just the same as
in the Schwarzchild black hole case. On the other hand, the
tangential magnification is sensitive to the caustic structure,
which can be seen more clearly if we plug the solution of the lens
equation (\ref{lensgamma}) for $\delta \psi$ into Eq. (\ref{D_O}).
In fact, we have
\begin{equation}
\mu_t= (-1)^k\frac{D_{OS}}{D_{LS}}
\frac{3\sqrt{3}(1+\epsilon_{k})\xi}{2D_{OL}\left(R_k
\xi^3+\delta\hat\phi_s \right)}, \label{muthat}
\end{equation}
where the $(-1)^k$ accounts for the parity of the image and $\xi$
must be determined solving Eq. (\ref{Eqimages}). The whole
dependence of the magnification on the black hole spin and the
observer declination is through the caustic semi-amplitude $R_k$,
where they appear in the usual combination $a\sin \vartheta_o$.

\subsection{Relativistic images around Sgr A*}

In this subsection we want to complement the discussion about the
detectability of relativistic images done in Paper I by some
additional considerations. Indeed there are many factors that
contrast the positive detection of relativistic images around Sgr
A*. The photons with the right incident direction for performing a
complete loop around a black hole and then reach the observer are
very few, because a slight perturbation in the incident trajectory
results in a very different outgoing direction. Moreover, during
their journey, photons may be scattered or absorbed by the
accreting matter surrounding the supermassive black hole. Finally,
the photons surviving up to the observer must be recognized and
distinguished from the noise coming from the environment.

Scattering and absorption from accreting matter are strongly
model-dependent and cannot be easily quantified without
non-trivial assumptions on the infalling plasma physics. We are
not going to face this problem here, since it demands an extensive
investigation beyond the purpose of this work.

On the other hand, our gravitational lensing analysis allows us to
give sharp answers on the brightness and spatial properties of the
images. In Paper I, we have suggested that the observed Low-Mass
X-ray Binaries (LMXB) orbiting around Sgr A* provide an ideal
population of sources for the gravitational lensing in the SDL
\cite{Muno}. Of course we need to resolve the shadow of Sgr A* in
order to identify relativistic images around it. This requires a
resolution of the order of the $\mu$as, which is just one step
beyond the limit reached in the radio band. In the X-ray band,
projects of space interferometry which could reach resolutions
even better than $\mu$as are under study (MAXIM,
http://maxim.gsfc.nasa.gov). When such projects will become
reality, a complete imaging of Sgr A* will be possible and the
relativistic images could be distinguished.

Apart from spatial resolution, which can be attained by realistic
future projects, in order to detect a signal in the X-ray band
from a relativistic image, we need a sufficient energy flux. With
an intrinsic luminosity $L_S\sim 2\times 10^{33}$ ergs s$^{-1}$ in
the band $2-10$ keV, emitted by a surface with radius $R_S=100$km,
LMXBs are as powerful sources as Sgr A* itself but enjoy a much
higher surface brightness \cite{Muno}. If one of these sources
crosses a relativistic caustic of order $k$, the angular area of
the resulting relativistic image is the original source area $\pi
R_S^2/D_{OS}^2$ multiplied by the magnification factor $\mu$. As
long as the source is inside the caustic, the magnification stays
higher than a minimum value corresponding to a source located at
the center of the caustic. The central magnification has been
calculated in Paper I and amounts to

\begin{equation}
\mu_c=\frac{D_{OS}^2}{D_{LS}^2D_{OL}^2}\frac{27\epsilon_k(1+\epsilon_k)}{4R_k},
\end{equation}
for each of the four relativistic images present when the source
is inside the caustic.

For a detector with collecting area $A_D$, the observed flux,
taking into account an absorption factor $\varepsilon=0.158$,
deduced from Ref. \cite{Bagan}, is thus
\begin{equation}
F_k=\varepsilon \frac{L_S}{4\pi R_S^2} \left(\mu_c \frac{\pi
R_S^2}{D_{OS}^2} \right)A_D.
\end{equation}

With $D_{OL}=8$kpc, $M_{BH}=4.3\times 10^6 M_\odot$ \cite{Bel} and
$D_{OS} \simeq D_{OL}$ (since $D_{LS} \ll D_{OL}$), we have
\begin{equation}
F_2=2.3\times 10^{-11} \mathrm{ergs ~ s^{-1}} \left(
\frac{D_{LS}}{100AU}\right)^{-2}  \left(
\frac{a}{0.02}\right)^{-2} \left( \frac{A_D}{100m^2} \right),
\end{equation}
for a source crossing the caustic of order $k=2$ and a black hole
spin $a=0.02$ \cite{LiuMel}. This flux is independent of the
source radius, as long as the source is much smaller than the
caustic extension, as in our case. We have considered a collecting
area $A_D=100m^2$ which might be realistically obtained by future
space detectors. The count rate for photons in the considered band
(with average energy 6 keV) is thus of the order of $2.4\times
10^{-3}$ s$^{-1}$, which is comparable to the counts usually
reported as positive detections by the Chandra satellite for faint
sources \cite{Bagan,Muno}. Of course, such a high value for the
count rate can only be achieved with a collecting area as large as
that we have considered here, which is roughly 100 times larger
than that of Chandra.

Sgr A* itself emits in the X-rays and provides a background noise
to the signal of a relativistic image. The image of an LMXB is
entirely contained within a single pixel of a hypothetic detector
where every pixel covers $1\mu$as$\times 1\mu$as of sky. We can
estimate the noise due to Sgr A* considering that its intrinsic
luminosity is of the same order as $L_S$\cite{Bagan,Muno}, but its
emitting region has a radius $R_{Sgr}$ of the order of 100
Schwarzschild radii. Then, every pixel is affected by a noise from
Sgr A* of the order of
\begin{equation}
F_{Sgr}=\varepsilon \frac{L_S}{4\pi R_{Sgr}^2} \omega_p^2 A_D,
\end{equation}
where $\omega_p$ is the size of the pixel. We thus have
\begin{equation}
F_{Sgr}=3.7\times 10^{-14} \mathrm{ergs ~ s^{-1}}\left(
\frac{\omega_p}{1\mu\mathrm{as}}\right)^{2} \left(
\frac{A_D}{100m^2} \right),
\end{equation}
which is roughly 600 times smaller than $F_2$. This proves that
the background from Sgr A* is indeed negligible for relativistic
images of order 2 if one has sufficient resolving power. It is
also important to stress that these estimates has been calculated
considering the minimum magnification $\mu_c$ for a source inside
a caustic. When the source is close to a fold or a cusp, the
brightness of the relativistic image can be sensibly higher.

We conclude this discussion mentioning that the brightness of
relativistic images of order 3 is $0.016F_2$, which allows a
marginal detection w.r.t. the noise by Sgr A*, while relativistic
images of higher order are too faint to be detected, at least for
the configuration examined here.

\section{Conclusions}

This paper completes the cycle of papers devoted to the study of
gravitational lensing by Kerr black holes in the Strong Deflection
Limit. After the first pioneering work of Ref. \cite{BozEq}, where
equatorial lensing was reduced to the same problem already solved
for spherically symmetric black holes \cite{Boz1}, in Ref.
\cite{BDLSS} we managed to make a complete analytical treatment of
Kerr lensing for equatorial observers, introducing a perturbative
expansion in the spin $a$. In this work we have extended that idea
to Kerr lensing with a generic observer. Though the strategy is
essentially unchanged, the introduction of a new parameter (the
inclination of the spin or equivalently the observer colatitude
$\vartheta_o$) has increased the difficulty of the derivation.
Nevertheless, our investigation has reached its objective: a
basically simple and accurate description of Kerr lensing
phenomenology with arbitrary observer position.

An essential summary of the main results obtained includes: the
shape of the shadow of the black hole (\ref{shadow}); the shape of
all critical curves (\ref{CritEllipse}); the shape and position of
the caustics (Eqs. (\ref{cauhattheta}) and (\ref{cauhatphi})); the
position of the images (Eqs. (\ref{2image1})-(\ref{2image2}) with
Eq. (\ref{Eqimages})) and their magnification (Eqs.
(\ref{magnrad}) with (\ref{eigenrad}) and (\ref{muthat})) for
sources close to a caustic.

To the second order in $a$, the shadow of the black hole and the
critical curves are ellipses slightly displaced from the black
hole position. The ellipticity is slightly higher in critical
curves than in the shadow. The caustics are displaced from the
optical axis and show the characteristic 4-cusped astroid shape
with the same extension in both directions. The caustic shrinks
back to a single point when the observer lies on the spin axis,
restoring the axial symmetry. There are two additional images when
the source is inside a caustic.

The fundamental fact that emerges is that all observables to the
lowest order are functions of $a \sin \vartheta_o$, which
represents the projection of the black hole spin on a plane
orthogonal to the line of sight. These observables include: the
shift and the ellipticities of the shadow and of critical curves;
the shift and the extension of the caustics; the position and the
magnification of the images.

The degeneracy between the absolute value of the spin and its
inclination on the line of sight can only be broken by
next-to-leading order terms in all observables. This has been
explicitly shown considering the shadow and critical curves
semi-axes and the caustic vertical shift. These are second order
contributions to zero-order quantities, thus requiring extremely
accurate measures, which may be very challenging. For example, if
the black hole spin is $a=0.1$, in order to break the degeneracy
we need a relative accuracy of order $a^2=0.01$ in the measures.

The most promising way to break the degeneracy is through higher
order corrections to the positions of the images. In fact, our
second order treatment is only sufficient to determine the
position angle of the images to zero order in $a$. Indeed the
first order corrections are likely to be at reach of future VLBI
observations, but unfortunately they require at least a third
order treatment of Kerr lensing in order to be determined
analytically. This could represent the main target for future
theoretical developments of our methodology.

Of course, if the black hole spin is close to the extremal value
$a=0.5$, the degeneracy breaking terms arising from higher orders
in $a$ grow to the same size as the lowest order contributions and
the problem would not be the degeneracy between $a$ and
$\vartheta_o$ but the correct theoretical interpretation of the
observations in a non-perturbative frame, in order to perform a
safe parameters extraction.

\begin{acknowledgments}
V.B. and G.S. acknowledge support for this work by MIUR through
PRIN 2004 ``Astroparticle Physics'' and by research fund of the
Salerno University. F.D.L.'s work was performed under the auspices
of the EU, which has provided financial support to the ``Dottorato
di Ricerca Internazionale in Fisica della Gravitazione ed
Astrofisica'' of the Salerno University, through ``Fondo Sociale
Europeo, Misura III.4''.
\end{acknowledgments}

\appendix

\section{Resolution of radial integrals}

This appendix reports the calculation of the radial integrals
appearing in the geodesics equations (\ref{Geod1}) and
(\ref{Geod2}). The double signs remind us that the integration
along the whole trajectory of the photon must be performed in such
a way that all pieces bounded by two consecutive inversion points
must sum up with the same sign \cite{Cha}. Gravitational lensing
trajectories have only one inversion point in $x_0$, the closest
approach distance. Thus we just have to sum the contributions due
to two branches (approach and departure). These two branches of
the photon trajectory are actually related by the time-reversal
symmetry, so that the results of the whole radial integrals are
just twice the contributions covering the departure branch.
Summing up, the radial integrals reduce to
\begin{eqnarray}
&& I_1=2\int\limits_{x_0}^{\infty} \frac{dx}{\sqrt{R}}
\label{I1}\\
&& I_2=2\int\limits_{x_0}^{\infty} \frac{x^{2}+a^{2}-a J}{\Delta
\sqrt{R}} dx. \label{I2},
\end{eqnarray}
where we have neglected the corrections due to the finiteness of
$D_{OL}$ and $D_{LS}$, thus extending the integration domain to
$+\infty$. The resolution by the SDL technique can be read from
the appendix A of Paper I, since the only change comes when we
replace $J$ and $Q$ by their new expressions containing $\mu_o$.
Thus we can directly jump to the results, which read
\begin{eqnarray}
& I_1=&- a_1 \log \delta+ b_1 \\
& I_2=&- a_2 \log \delta+ b_2.
\end{eqnarray}

The coefficients expanded to second order in $a$ are
\begin{eqnarray}
&& \!\!\!\!\!\!\!\! \!\!\!\!\!\!\!\! \!\!\!\!\!\!\!\! a_1 = \frac{4}{3\sqrt{3}}+\frac{16}{27}a \hat\xi+\frac{8}{81\sqrt{3}}a^2(7+4\mu_o^2+5\hat\xi^2) \\
&& \!\!\!\!\!\!\!\! \!\!\!\!\!\!\!\! \!\!\!\!\!\!\!\!
b_1=a_1\log[12(2-\sqrt{3})]
-\frac{8}{81}(5\sqrt{3}-6)a^2(1-\hat\xi^2)
\end{eqnarray}
\begin{eqnarray}
& a_2 &=  \frac{4}{\sqrt{3}}+\frac{8}{3}a\hat\xi+\frac{8}{9\sqrt{3}}a^2(3+2 \mu_o^2+5\hat\xi^2) \\
& b_2 &=  a_2 \log \left[4\sqrt{3}(2\sqrt{3}-3)^{1+\sqrt{3}}
\right]\nonumber
\\ && -\frac{8}{9} a \hat\xi [9-2\sqrt{3}+3\sqrt{3}\log(2\sqrt{3}-3)]
 \nonumber \\ &&
+\frac{4}{27}a^2 \left\{26\sqrt{3}-16-2\sqrt{3}\log(3)
+8\mu_o^2-12\sqrt{3}\mu_o^2 \right. \nonumber \\ && \left.
+12(3-\mu_o^2)\log(2\sqrt{3}-3) -\hat\xi^2[38\sqrt{3}-20 \right.
\nonumber \\ && \left. +5(1+2\sqrt{3})\log(3)+30\log(2-\sqrt{3})]
\right\}
\end{eqnarray}
with $\hat\xi=\xi\sqrt{1-\mu_o^2}$.

\section{Resolution of angular integrals}

This appendix is devoted to the resolution of the angular
integrals
\begin{eqnarray}
&& J_1=\pm \int \frac{1}{\sqrt{\Theta}} d \vartheta \\
&& J_2=\pm \int \frac{\csc^2\vartheta}{\sqrt{\Theta}} d \vartheta
.
\end{eqnarray}

Introducing the variable $\mu=\cos \vartheta$, the two integrals
become
\begin{eqnarray}
&& J_1=\pm \int \frac{1}{\sqrt{\Theta_\mu}} d \mu \label{J1mu}\\
&& J_2=\pm \int \frac{1}{(1-\mu^2)\sqrt{\Theta_\mu}} d \mu
,\label{J2mu}
\end{eqnarray}
where
\begin{eqnarray}
& \Theta_\mu=&a^2(\mu_-^2+\mu^2)(\mu_+^2-\mu^2) \\
& \mu_\pm^2=&\frac{\sqrt{b_{JQ}^2+4a^2Q_m}\pm b_{JQ}}{2 a^2} \label{mupm}\\
& b_{JQ}=& a^2-J_m^2-Q_m,
\end{eqnarray}
and we have already replaced $J$ and $Q$ with $J_m$ and $Q_m$,
coherently with the fact that we only retain terms that are
logarithmically diverging or constant in the approach parameter
$\delta$ (or equivalently in the separation from the shadow
$\epsilon$).

$\Theta_\mu$ has two zeros in $\mu=\pm \mu_+$. Then the photon
performs symmetric oscillations of amplitude $\mu_+$ w.r.t. the
equatorial plane. It is useful to write the explicit expressions
of $\mu_+$ and $\mu_-$ in terms of the spin $a$ and the position
parameter $\xi$. Using Eqs. (\ref{Jma2})-(\ref{Qma2}) in Eq.
(\ref{mupm}) and expanding to the second order in $a$, we find
\begin{eqnarray}
& \mu_{+}=&\sqrt{1-\hat\xi^2}\left[ 1+a A_
+ + \frac{1}{2}a^2 A_+^2 \right]\\
& \mu_{-}=&\frac{3 \sqrt{3}}{2 a}-2 \hat\xi -4
a\frac{(\mu_o^2+\hat\xi^2)}{3 \sqrt{3}} \nonumber \\
&&+\frac{4}{27}a^2 \hat\xi (3-10\mu_o^2-8\hat\xi^2)
\end{eqnarray}
where
\begin{equation}
A_+=\frac{2 \hat\xi(1-\mu_o^2)(1-\xi^2)}{3\sqrt{3}(1-\hat\xi^2)}
\end{equation}
In a first approximation, the oscillation amplitude $\mu_+$ is
$\sqrt{1-\hat\xi^2}$, plus corrections due to the black hole spin.
Note that the minimal amplitude of the oscillations is obtained
for $\xi=\pm 1$, which gives $\mu_+=|\mu_o|$. Purely equatorial
trajectories with $\mu_+=0$ are involved in gravitational lensing
only if the observer itself lies on the equatorial plane. On the
other hand, polar photons ($\xi=0$) perform oscillations with
maximal amplitude $\mu_+=1$, touching the poles of the black hole.

Now it is convenient to introduce a new integration variable
$z=\mu/\mu_+$, which allows to eliminate the dependence on $a$ in
the integration extrema. The integrals become
\begin{eqnarray}
&& J_1=\pm \int \frac{1}{\sqrt{\Theta_z}} d z \label{J1z}\\
&& J_2=\pm \int \frac{1}{(1-\mu_+^2z^2)\sqrt{\Theta_z}} d z
,\label{J2z}
\end{eqnarray}
with
\begin{equation}
\Theta_z=a^2(\mu_-^2+\mu_+^2z^2)(1-z^2).
\end{equation}

In order to perform the angular integrals, it is wise to expand
the integrands to second order in $a$ and then integrate. The
primitive functions read
\begin{eqnarray}
& F_{J_1}(z) &=\frac{2}{3\sqrt{3}}\arcsin (z) \left[1+\frac{4}{3\sqrt{3}}a\hat\xi \right. \nonumber \\
&& \left. -\frac{1}{27}a^2 (1-8\mu_o^2-25\hat\xi^2) \right]  \nonumber \\
&& +\frac{2}{81\sqrt{3}}a^2 (1-\hat\xi^2)z\sqrt{1-z^2} \label{FJ1}
\end{eqnarray}
\begin{eqnarray}
& F_{J_2}(z) &= \frac{2}{3\sqrt{3}\hat\xi}\arctan \left[\frac{z
\hat\xi}{\sqrt{1-z^2}} \right] \left\{1 \right. \nonumber \\ &&
\left. + \frac{2}{3\sqrt{3}\hat\xi}a (1-\mu_o^2+\hat\xi^2) +
\frac{2}{27\hat\xi^2}a^2 \left[ 3(1+\mu_o^4-\hat\xi^2) \right.
\right. \nonumber \\ && \left. \left.
+11\hat\xi^4-6\mu_o^2(1-\hat\xi^2)\right] \right\}
+\frac{4}{81\sqrt{3}}a^2 \arcsin (z) \nonumber \\ &&
-\frac{4}{27\hat\xi}  \frac{z
\sqrt{1-z^2}(1-\mu_o^2-\hat\xi^2)}{1-(1-\hat\xi^2)z^2}\left\{ a \right. \nonumber \\
&& \left. + \frac{a^2}{3\sqrt{3}\hat\xi\left[ 1-(1-\hat\xi^2)z^2
\right]} \left[ 3(1-\mu_o^2)+\hat\xi^2 \right. \right. \nonumber \\
&& \left. \left.
-z^2\left(3-4\hat\xi^2+\hat\xi^4-\mu_o^2(3-5\hat\xi^2) \right)
\right] \right\} \label{FJ2}
\end{eqnarray}

Similarly to radial integrals, the angular integrals appear with
double signs reminding that they must be performed piece by piece
between any two consecutive inversion points and all contributions
must be summed with the same sign \cite{Cha}. The integration
covers the whole trajectory of the photon, which may perform
several oscillations around the equatorial plane. The integration
must start from the source position $z_s\equiv \mu_s/\mu_+$ and
must end at the observer position $z_o \equiv \mu_o/\mu_+$. Let us
indicate by $m$ the number of inversion points in the polar motion
touched by the photon. Still we must consider two possibilities
depending on the direction taken by the photon starting from
$z_s$. In fact, we may have a trajectory in which $z$ is initially
either growing or decreasing. In the first case, the first pieces
of the angular integrals cover the domain $[z_s,1]$. After that,
we have $m-1$ integrals covering the whole domain $[-1,1]$. All
these integrals must be taken with the same sign so that they
always sum up. Finally, if $m$ is even, the photon reaches $z_o$
with growing $z$ and the last piece covers the domain $[-1,z_o]$,
otherwise $z$ is finally decreasing and the domain is $[z_o,1]$.
The total angular integrals are thus given by the sum of all these
contributions covering the domains just described. Exploiting the
primitive functions (\ref{FJ1}) and (\ref{FJ2}), we can express
each integral as (in the following, $i$ takes the values 1 or 2)
\begin{eqnarray}
&J_i&=F_{J_i}(1)-F_{J_i}(z_s)
+(m-1)\left[F_{J_i}(1)-F_{J_i}(-1)\right] \nonumber \\ && +
F_{J_i}(z_o)-F_{J_i}(-1)
\end{eqnarray}
for $m$ even and
\begin{eqnarray}
&J_i&=F_{J_i}(1)-F_{J_i}(z_s)
+(m-1)\left[F_{J_i}(1)-F_{J_i}(-1)\right] \nonumber \\ && +
F_{J_i}(1)-F_{J_i}(z_o)
\end{eqnarray}
for $m$ odd.

Noting that both primitives are odd functions of $z$, we have
$F_{J_i}(-1)=-F_{J_i}(1)$ and we can express the angular integrals
in the compact form
\begin{equation}
J_i=\mp \left[F_{J_i}(z_s) - (-1)^m F_{J_i}(z_o) \right] +2m
F_{J_i}(1).
\end{equation}
The $(-1)^m$ ensures that the sign of the $z_o$-term is the same
as the $z_s$-term if the number of inversions is odd and is
opposite if $m$ is even. We have also introduced a double sign to
take into account the possibility that $z$ is initially decreasing
from the starting value $z_s$.

For future reference, we also write the explicit values of
$F_{J_i}(1)$
\begin{eqnarray}
&F_{J_1}(1)&=
\frac{\pi}{3\sqrt{3}}\left[1+\frac{4a\hat\xi}{3\sqrt{3}} \right.
\nonumber \\ && \left.
-\frac{a^2(1-8\mu_o^2-25\hat\xi^2)}{27} \right] \\
&F_{J_2}(1)&= \frac{\pi}{3\sqrt{3}\hat\xi} \left\{1+ \frac{2a}{3\sqrt{3}\hat\xi} (1-\mu_o^2+\hat\xi^2) \right. \nonumber \\
&& + \frac{2a^2}{27\hat\xi^2}
\left[ 3(1+\mu_o^4-\hat\xi^2)+\hat\xi^3+11\hat\xi^4 \right. \nonumber \\
&&
 \left. \left. -6\mu_o^2(1-\hat\xi^2) \right] \right\}.
\end{eqnarray}

\section{Second order contributions to the lens equation}

In this appendix we report the expressions for $\delta
\mu_s^{(2)}$ and $\delta \phi_s^{(2)}$, which must be added to
Eqs. (\ref{Eqmu1}) and (\ref{Eqphi1}) to obtain the second order
lens equation. They read

\begin{eqnarray}
&\delta \mu_s^{(2)}&=\mp(-1)^m\frac{\sqrt{1-\hat\xi^2}}{54}
\left\{6(\hat\xi^2-1)\cos \psi_1 \right. \nonumber \\&&
+\left[(1+3\hat\xi^2)\cos \psi_o \right. \nonumber \\&&
\left.\left.
 + (1-\hat\xi^2) \cos (2\psi_1+\psi_o)
\right] \sin \psi \right\}
\end{eqnarray}
\begin{eqnarray}
&\delta \phi_s^{(2)}&= \frac{1-\hat\xi^2}{27\hat\xi^2}
\cos^2\psi_o\left[ 1-21\hat\xi^2+(1-\hat\xi^2)\cos 2\psi_o \right]
\cdot \nonumber \\
 && \cdot \left[ \arctan(\hat\xi \tan \psi_o)-
\arctan(\hat\xi \tan \psi_1) \right] \nonumber \\ &&
-\frac{2+(1-\hat\xi^2)\cos 2\psi_1} {9(\cos^2\psi_1+\hat\xi^2
\sin^2
\psi_1)}\hat\xi \psi_n \nonumber \\
 &&
+\frac{1}{864\hat\xi(\cos^2\psi_1+\hat\xi^2 \sin^2 \psi_1)^2}
\sum\limits_{i=0}^3 p_i \hat\xi^{2i},
\end{eqnarray}
where $\psi_1=\psi+\psi_o$, $\psi_n=5\psi+8\sqrt{3}-20$ and
\begin{eqnarray}
&p_0&=64\cos^3\psi_1\cos^3 \psi_o \sin \psi \\ &p_1&=384 \cos
\psi_1+2(7+4\cos 2\psi_o+\cos 4\psi_o)\sin 2\psi_1 \nonumber
\\ && - (11+20\cos 2 \psi_o+5 \cos 4\psi_o)\sin 4\psi_1 \nonumber
\\ && +4\left[384+(14\sin 2\psi_o+5 \sin 4\psi_o)\cos^2\psi_1
\right]\cos 2 \psi_1 \nonumber \\
&&+96(12-\cos^2\psi_1\cos^3\psi_o \sin \psi_o) \\ &  p_2 & =
768(1-\cos 4\psi_1)-8 (9+\cos 2 \psi_o)\cos^2 \psi_o \sin 2 \psi_1
\nonumber \\ && + (13+28\cos 2\psi_o+7\cos 4 \psi_o) \sin 4 \psi_1
\nonumber \\ && +8(9-11\cos 2 \psi_1) \cos^2\psi_1\sin 2 \psi_o
\nonumber \\ && +4(9+7\cos 2 \psi_1) \sin^2 \psi_1 \sin 4 \psi_o
\\
& p_3 & = 384(3+\cos 4 \psi_1)-(5+3\cos 4\psi_o)\sin 4 \psi_1
\nonumber \\ && +2(9-\cos 4\psi_o+24 \cos 2\psi_o \sin^2\psi_1)
\nonumber \\ && -20(\sin^2
2\psi_1 \sin 2 \psi_o + \sin^2 \psi_1\sin 4 \psi_o) \nonumber \\
&& -12 (128+\sin^2\psi_1 \sin 4 \psi_o) \cos 2\psi_1.
\end{eqnarray}

\end{document}